\documentclass[twocolumn,showpacs,amsfonts,aps,prc,nofootinbib,floatfix]{revtex4}
\usepackage{amsmath,amssymb,amsfonts}
\usepackage{appendix}
\usepackage{graphics,graphicx}
\usepackage{algorithm}
\usepackage{algpseudocode}
\usepackage[normalem]{ulem}
\usepackage{color}
\usepackage{float}

\newcommand\eqcom[1]{\stackrel{#1}{=}}

\begin{document}

%%%%%%%%%%%%%%%%%%%%%%%%%%%%
%% header info
%%%%%%%%%%%%%%%%%%%%%%%%%%%%
	\title{Hanbury Brown-Twiss correlation functions and radii
			\\ from event-by-event hydrodynamics}
	\author{Christopher Plumberg}
	\author{Ulrich Heinz}
	\affiliation{Department of Physics, The Ohio State University,
					Columbus, OH 43210-1117, USA}

	%%%%%%%%%%%%%%%%%%%%%%%%%%%%
	%% abstract here
	%%%%%%%%%%%%%%%%%%%%%%%%%%%%
	\begin{abstract}
		\textbf{HoTCoffeeh} (\textbf{H}anbury Br\textbf{o}wn-\textbf{T}wiss \textbf{Co}rrelation
		\textbf{f}unctions and radii \textbf{f}rom \textbf{e}vent-by-\textbf{e}vent
		\textbf{h}ydrodynamics) is a new computational tool which determines Hanbury Brown-Twiss
		(HBT) charged pion ($\pi^+$) correlation functions and radii for event-by-event (EBE) 
		hydrodynamics with fluctuating initial conditions in terms of Cooper-Frye integrals, 
		including resonance decay contributions.
		In this paper, we review the basic formalism for computing the HBT correlation functions
		and radii with resonance decay contributions included, and discuss our implementation of
		this formalism in the form of HoTCoffeeh.  This tool may be easily integrated with other
		numerical packages (e.g., \cite{Shen:2014vra}) for the purpose of simulating the evolution
		of heavy-ion collisions and thereby extracting predictions for heavy-ion observables.
	\end{abstract}
	%%%%%%%%%%%%%%%%%%%%%%%%%%%%
	%%%%%%%%%%%%%%%%%%%%%%%%%%%%

	%%%%%%%%%%%%%%%%%%%%%%%%%%%%
	%% miscellanea
	%%%%%%%%%%%%%%%%%%%%%%%%%%%%
	\date{\today}

	\maketitle
	
	%%%%%%%%%%%%%%%%%%%%%%%%%%%%%%%%%%%%%%%%%%%%%%%%%%%
	%% set user-defined commands, shortcuts, etc. here
	%%%%%%%%%%%%%%%%%%%%%%%%%%%%&&&&&&&&&&&&&&&&&&&&&&&
	\def\l{\left}
	\def\r{\right}
	\def\la{\langle}
	\def\ra{\rangle}
	\def\dla{\la\!\la}
	\def\dra{\ra\!\ra}
	\def\BIGGL{\scalebox{1.25}{\Bigg[}}
	\def\BIGGR{\scalebox{1.25}{\Bigg]}}
	\def\ev{\mathrm{ev}}
	\def\e{\mathrm{e}}
	\def\eq{{\,=\,}}
	\def\mcO{{\mathcal{O}}}
	\def\mcP{{\mathcal{P}}}
	\def\mco{o}
	\def\Mmax{M_{\mathrm{max}}}
	\def\mrP{{\mathrm{phys}}}
	\def\mrM{{\mathrm{meas}}}
	\def\nev{N_{\mathrm{ev}}}
	\def\mfp{\mathrm{mfp}}
	\def\ebsT{\l( \eta/s \r)(T)}
	\def\QCD{\mathrm{QCD}}
	\def\dir{\mathrm{dir}}
	
	\newcommand{\evavg}[1]{\l< #1 \r>_{\ev}}
	\newcommand{\Savg}[1]{\bigl\langle #1 \bigr\rangle_S} % for average
	\newcommand{\avg}[1]{\bigl\langle #1\bigr\rangle} % for average
	\newcommand{\twopart}{\frac{dN}{d^3p_1d^3p_2}}
	\newcommand{\onepart}[1]{\frac{dN}{d^3 #1}}
	\newcommand{\re}{\mbox{Re}}
	\newcommand{\im}{\mbox{Im}}
	\newcommand{\sh}{\mbox{sh}}
	\newcommand{\ch}{\mbox{ch}}
	\newcommand{\erf}{\mbox{erf}}
	\newcommand{\azavg}[1]{\left< #1 \right>_{\Phi_K}} % for average
	\newcommand{\expect}[1]{\mbox{E} \left[ #1 \right]}
	\newcommand{\expectn}[2]{\mbox{E}_{#2}\left[ #1 \right]}
	\newcommand{\var}[1]{\mbox{Var} \left[ #1 \right]}
	\newcommand{\cov}[2]{\mbox{Cov} \left( #1, #2 \right)}

	%%%%%%%%%%%%%%%%%%%%%%%%%%%%
	%% Paper begins here
	%%%%%%%%%%%%%%%%%%%%%%%%%%%%
	\section{Introduction}
	\label{Sec1}
	Hanbury Brown-Twiss (HBT) interferometry (also known as \textit{femtoscopy}) is a technique which has been used successfully over the past several decades of heavy-ion physics to probe the spatio-temporal and dynamical properties of freeze-out surfaces in relativistic heavy-ion collisions.  The observables derived from HBT interferometry, known as the HBT radii, thus provide a window into the shapes and sizes of these collisions.  Recently, the notion of studying the HBT radii on a collision-by-collision (or \textit{event-by-event}) basis has been raised \cite{Plumberg:2015eia, Plumberg:2015mxa, Plumberg:2015aaw}.  This has motivated two distinct questions: (1), whether event-by-event distributions of HBT radii (or \textit{HBT distributions}) are even experimentally accessible, and (2), if they are, what information they might contain about the properties of heavy-ion collisions. The studies presented in Refs.~\cite{Plumberg:2015eia, Plumberg:2015mxa, Plumberg:2015aaw} have answered the first question in the affirmative, and have addressed the second question by demonstrating that experimental measurements of the statistical moments of HBT distributions could potentially yield sensitivity to other interesting quantities, such as the value (and temperature dependence) of the specific shear viscosity $\eta/s$ in the quark-gluon plasma (QGP).  Probing HBT distributions experimentally may therefore yield valuable insights into the properties of relativistic heavy-ion collisions.
	
	One of the most successful theoretical and phenomenological approaches to date for understanding the properties of heavy-ion collisions involves numerically simulating the various stages of their evolution on an event-by-event basis, and using these simulations to make predictions which can be compared with experimental measurements of event-by-event heavy-ion observables.  In particular, a great deal of attention has been paid in this regard to event-by-event fluctuations of observables related to radial flow ($\avg{p_T}$) \cite{Abelev:2014ckr, Mazeliauskas:2015efa}, anisotropic flow ($v_n$) \cite{Alver:2007qw, Jia:2013tja, Aad:2013xma}, total multiplicity ($N^{\mathrm{ch}}$) \cite{Baym:1999up, Aguiar:2001ac, Aggarwal:2001aa}, and so on.  To extend the successes of this event-by-event hydrodynamic paradigm to include the HBT radii, then, clearly requires the ability to simulate HBT analyses on an event-by-event basis.
	
	An essential component of any HBT analysis, whether experimental or theoretical, involves properly accounting for the presence of resonance decay contributions to the final pion yields.  In the case of experimental HBT, these resonance decay contributions are a contaminating factor which can never be completely eliminated.  Theoretically, on the other hand, the resonance decay contributions must be computed separately, in addition to the thermally produced (or ``directly emitted") pions of interest.  Since the effects of resonances cannot be completely separated experimentally from the effects of direct pion emission, an apples-to-apples comparison between theory and experiment therefore requires theoretical analyses to compute the HBT radii with the resonance decay contributions included.  Thus, before the experimental accessibility of HBT distributions (and their connections to other aspects of heavy-ion physics) can be systematically explored from the perspective of hydrodynamics, one must first be able to compute the HBT radii on an event-by-event basis, with all relevant resonance decay contributions included.
	
	In the context of numerical simulations of heavy-ion collisions, such as those considered here, there are at least two different possible approaches to accomplishing this \cite{Qiu:2013wca}.  The first involves terminating the hydrodynamic evolution prior to kinetic freeze-out, converting the entire system to a collection of interacting hadrons whose initial distributions (at the point of conversion) are sampled from  Cooper-Frye distributions \cite{Cooper:1974mv}, and allowing those hadrons to scatter microscopically until all interactions cease because the matter has become too dilute.  Such an approach is typically called a ``hybrid" approach \cite{Petersen:2008dd}.  The second approach involves applying a hydrodynamic description of the system all the way until the entire system has reached kinetic freeze-out, implemented as a sharp transition to free-streaming particles on the so-called freeze-out hypersurface $\Sigma_\mathrm{f}$.  This approach is often termed a ``purely hydrodynamic" approach.  The hybrid approach has the advantage of describing the actual physical situation in heavy-ion collisions more realistically: two-particle correlations are always probed \emph{experimentally} with only a finite number of particles per event. This means that, in order to obtain good statistical precision, the hadronic ``afterburner'' must be run multiple times on a single hydrodynamic event. In the purely hydrodynamic approach the two-particle correlations (and therefore the associated HBT radii) are computed as Cooper-Frye integrals \cite{Cooper:1974mv} of the phase-space-distribution function over the freeze-out surface which provides results without statistical uncertainties (in effect implementing the assumption that each event emits an infinite number of particles).   In this paper, we adopt the purely hydrodynamic approach.
	
	As we discuss below, this problem is computationally complex, especially when the ${\sim\,}1700$ pion producing decay channels of the ${\sim\,}340$ species of resonances with masses below 2\,GeV created in the collision are included, as discussed in Ref.~\cite{Qiu:2012tm}.  
In fact, this complexity is the primary reason that the studies \cite{Plumberg:2015eia, Plumberg:2015mxa, Plumberg:2015aaw} omitted the resonance decay contributions from their analyses.  In this paper, we introduce a code designed to address the challenge of efficiently computing HBT correlation functions and radii (with all resonance decay effects included) from pure hydrodynamics on an event-by-event basis, thereby allowing a systematic exploration of HBT distributions in heavy-ion collisions to be performed within a reasonable timeframe.

	This paper is organized as follows.  In Sec.~\ref{Sec2b}, we will define the two-particle correlation function, and Sec.~\ref{Sec2c} will show how the HBT radii are extracted from the correlation function.  In order to interpret these radii in terms of spatio-temporal structure of the emitting source, we will show in Sec.~\ref{Sec2d1} how they may be computed directly from the emission function which defines the emission probability along the freeze-out surface.  In particular, we will show that the HBT radii may be determined from the Fourier-transform of the emission function, and will use this feature to show how they can be related directly to the emission function.  The emission function, in turn, will generally receive contributions from particles emitted directly by the source, as well as from particles which are decay products of other directly emitted particles.  In Sec.~\ref{Sec2d2}, we will show how to include resonance decay effects in the definition of the emission function, thus allowing us to explore the corresponding effects induced by these resonance decay contributions to the extracted HBT radii.  Finally, in Sec.~\ref{Sec3}, we will present some of the numerical results obtained using our code, and show how these results compare with the results of previous theoretical HBT analyses.

	%%%%%%%%%%%%%%%%%%%%%%%%%%%%%%%%%%%%%%%%%%%%%%%%%%%%%%%
	\section{HBT Formalism}
	\label{Sec2}
	%%%%%%%%%%%%%%%%%%%%%%%%%%%%%%%%%%%%%%%%%%%%%%%%%%%%%%%
	The formalism needed for the application of HBT interferometry \cite{HanburyBrown:1954amm,Brown:1956zza,HanburyBrown:1956bqd} to relativistic heavy-ion collisions (a.k.a. femtoscopy) is well established, and the reader is referred to Refs.~\cite{Gyulassy:1979yi,Heinz:1996bs,Heinz:1999rw,Wiedemann:1999qn,Lisa:2005dd,Lisa:2008gf} for reviews. For a self-contained presentation we review here briefly only the most essential definitions and relations.
	
	%%%%%%%%%%%%%%%%%%%%%%%%%%%%%%%%%%%%%%%%%%%%%%%%%%%%%%%
	\subsection{Correlation functions}
	\label{Sec2a}
	%%%%%%%%%%%%%%%%%%%%%%%%%%%%%%%%%%%%%%%%%%%%%%%%%%%%%%%
	
	The fundamental object of HBT interferometry is the two-particle momentum correlation function among pairs of particles from a single collision event,
	\begin{equation}
	C(\vec{p}_1, \vec{p}_2) \equiv \frac{ E_{p_1} E_{p_2} \frac{d^6 N}{d^3 p_1 d^3 p_2}}{ \l( E_{p_1} \frac{d^3 N}{d^3 p_1}\r) \l(  E_{p_2} \frac{d^3 N}{d^3 p_2} \r) }. 
	\label{corrfuncSEdefn}
	\end{equation} 
	Even at Large Hadron Collider (LHC) energies the number of particle pairs emitted from a single collision is, however, not large enough to measure this correlation function as a function of all six momentum components with adequate statistical precision. Instead of Eq.~(\ref{corrfuncSEdefn}) experiments therefore measure the ensemble-averaged correlation function
	\begin{equation}
	C_{\mathrm{avg}}(\vec{p}_1, \vec{p}_2) \equiv \frac{ \evavg{E_{p_1} E_{p_2} \frac{d^6 N}{d^3 p_1 d^3 p_2} }}{ \evavg{ E_{p_1} \frac{d^3 N}{d^3 p_1}} \evavg{ E_{p_2} \frac{d^3 N}{d^3 p_2} } }\label{corrfuncEAdefn}
	\end{equation} 
	where the signal pairs in the numerator and the product of single-particle distributions in the denominator (obtained from collecting uncorrelated pairs from mixed events \cite{Lisa:2005dd}\footnote{%
	In practice a much larger number of mixed-event pairs than signal pairs are generated, to 
	minimize the statistical uncertainty of the denominator of (\ref{corrfuncSEdefn}). For simplicity, we ignore
	this and the associated normalization factor of the mixed-event pairs.}%
) are averaged over sufficiently large sets of collision events with identical event characteristics:
	\begin{equation}
		\evavg{\cdots} \equiv \frac{1}{\nev} \sum^{\nev}_{k=1} \l( \cdots \r)_k.
		\label{EvAvgDefn}
	\end{equation}
	%

	%%%%%%%%%%%%%%%%%%%%%%%%%%%%%%%%%%%%%%%%%%%%%%%%%%%%%%%
	\subsection{HBT radii}
	\label{Sec2b}
	%%%%%%%%%%%%%%%%%%%%%%%%%%%%%%%%%%%%%%%%%%%%%%%%%%%%%%%
	The correlation function \eqref{corrfuncEAdefn}, after being corrected for final-state interactions, exhibits an enhancement for bosons (or a depletion for fermions) near $\vec{q}=0$ which, for spatially Gaussian source functions, can be fitted to a Gaussian in $q$:\footnote{
	The form of Eq. \eqref{corrfunc_functional_defn} neglects the effects of final-state interactions
	such as the long-range Coulomb repulsion which is inevitably present between electrically charged
	pairs of identical particles (e.g., $\pi^+$).  These interactions lead to a \emph{reduction} of
	particle pairs at small $\vec{q}$.  Fortunately, it is possible to account for these interactions
	in the experimental construction of the correlation function \cite{Adams:2004yc}.  We here 
	assume that the ensemble-averaged correlation function \eqref{corrfuncEAdefn} represents 
	such a Coulomb-corrected correlation function, and focus on the correlation effects caused by
	quantum statistics.
	}
	\begin{equation}
	C_{\mathrm{fit}}(\vec{q}, \vec{K}) \equiv 1 + \lambda(\vec{K}) \exp \l(-\sum_{i,j = o, s, l} R^2_{ij}(\vec{K}) q_i q_j \r). \label{corrfunc_functional_defn}
	\end{equation}
	Here we introduced the relative momentum $q^{\mu} = p_1^{\mu}{-}p_2^{\mu}$ between the two particles and their pair momentum $K^\mu = \frac{1}{2}\l( p_1^{\mu}{+}p_2^{\mu} \r)$, where $p_i^\mu$ are on-shell four-momenta with $p_1^2=p_2^2=m^2$ ($m$ being the mass of the (identical) particles whose correlation is measured). Due to this on-shell constraint the four-vectors $q$ and $K$ are orthogonal:
	\begin{equation}
		q \cdot K = 0. 
	\label{On_shell_condition_for_pions}
	\end{equation}
	The set of parameters $R^2_{ij}(\vec{K})$ in the exponent of Eq.~\eqref{corrfunc_functional_defn} are known as the HBT radius parameters (or ``HBT radii'' in short). They may be thought of as length scales characterizing the ``homogeneity regions" within the emitting source from which particle pairs with pair momentum $\vec{K}$ \cite{Makhlin:1987gm} are emitted. The Gaussian parametrization \eqref{corrfunc_functional_defn} is exact for emission functions with a Gaussian space-time structure. It is adequate even for non-Gaussian sources as long as their deviations from Gaussian structure are generated by additional length scales that are very different from the source radii.  Such additional length scales may be generated, for instance, by extremely long-lived resonances (e.g., the $\eta'$ meson) whose decays contribute to the final pion yield but are spread over a much larger region than the directly emitted pions.  These long-lived resonances lead to a sharp peak in the correlation function near $\vec{q}=0$ which is unresolvably narrow from an experimental standpoint. The experimental correlation function thus features only the much wider (in $q$) structure associated with directly emitted pions and those from short-lived resonance decays \cite{Wiedemann:1996ig,Heinz:1999rw}, while the contribution from long-lived resonances is invisible in the correlation function and thereby apparently reduces its value at $\vec{q} = 0$. This reduction effect is accounted for by the ``intercept parameter" $\lambda ( \vec{K} )$ in the functional form \eqref{corrfunc_functional_defn}.  
	
\begin{figure}[!htbp]
	 \centering
	 %The actual figure
	 \includegraphics[width=0.9\linewidth]{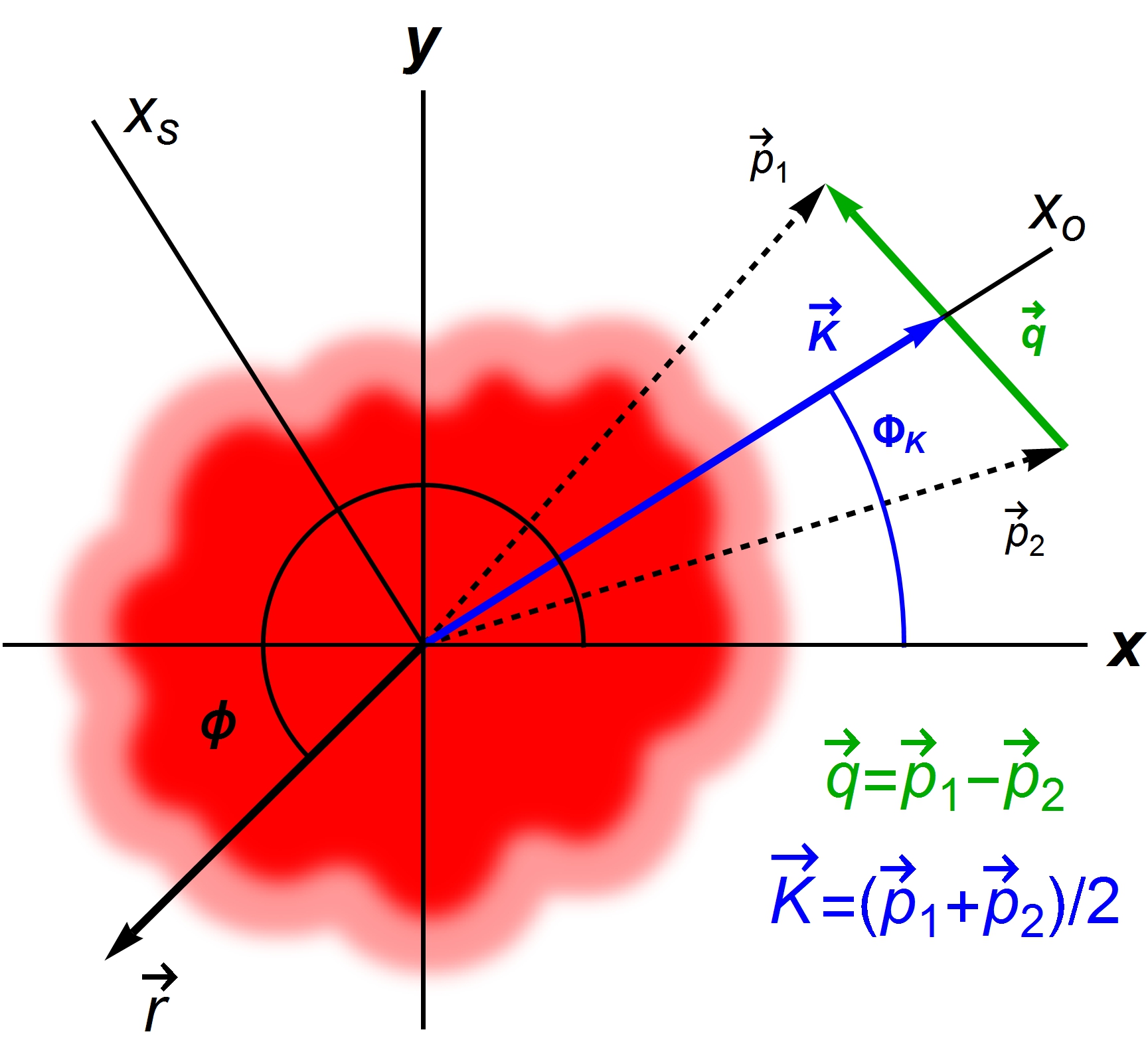}
	 \caption{(Color online)
	 The out-side-long ($osl$) coordinate system used for defining the HBT radii.  Here, $\vec{r}$, $\vec{p}_{1,2}$, $\vec{q}$ and $\vec{K}$ are seen projected into the transverse plane, so that the transverse component of $\vec{K}$ makes an angle $\Phi_K$ with the $x$-axis, defined to point in the direction of the impact parameter $\vec{b}$ ( or a proxy for it, such as the elliptic flow angle $\Psi_2$).  The longitudinal direction (i.e., the $z$-direction) is defined to point out of the page.  
	  \label{fig:oslcoords_with_r} 
	  }
	\end{figure}

	We refer to the extraction of HBT radii by a fit of the data with Eq.~(\ref{corrfunc_functional_defn}) as the Gaussian fit (or GF) method. In the sum in the exponent, $o,s,l$ denote the outward, sideward and longitudinal directions, respectively. These form a Cartesian coordinate system, with $l$ pointing along the beam (or $z$-) direction and $(o,s)$ spanning the transverse plane. The outward direction points along the transverse pair momentum $\vec{K}_T$ and forms an azimuthal angle $\Phi_K$ with the impact parameter $\vec{b}$ defining the $x$-axis (see Fig.~\ref{fig:oslcoords_with_r}).
	
	%%%%%%%%%%%%%%%%%%%%%%%%%%%%%%%%%%%%%%%%%
	%%%%%%%%%%%%%%%%%%%%%%%%%%%%%%%%%%%%%%%%%
	%%%%%%%%%%%%%%%%%%%%%%%%%%%%%%%%%%%%%%%%%
	%%%%%%%%%%%%%%%%%%%%%%%%%%%%%%%%%%%%%%%%%

	%%%%%%%%%%%%%%%%%%%%%%%%%%%%%%%%%%%%%%%%%%%%%%%%%%%%%%%
	\subsection{The emission function}
	\label{Sec2c}
	%%%%%%%%%%%%%%%%%%%%%%%%%%%%%%%%%%%%%%%%%%%%%%%%%%%%%%%
	Particle emission is characterized by the emission function $S(x,p)$, a single-particle Wigner function describing the phase-space distribution of the emitted particles. If averaged over phase-space volumes $\gg h^3$ it is positive definite and describes the probability for emitting a particle with momentum $p$ from point $x$. The emission function corresponding to the sudden freeze-out of a hydrodynamic fluid on a freeze-out surface $\Sigma_\mathrm{f}$ is a classical phase-space distribution and given by the Cooper-Frye prescription \cite{Cooper:1974mv,Schlei:1992jj,Chapman:1994xa}
	\begin{eqnarray}
		S(x,p) &=& \frac{1}{(2\pi)^3} \int_{\Sigma_\mathrm{f}} p{\,\cdot\,}d^3 \sigma(y)\, \delta^4 (x{-}y)\, f(y,p)\,,
	\label{cooper_frye_defn1}
	\\
	f(x,p) 	&=& f_0 \l(x,p\r) + \delta f \l(x,p\r)\nonumber\\
			&=& \frac{1}{e^{(p \cdot u{-} \mu)/T}{-}1}{+}\frac{\chi(p^2) p^{\mu} p^{\nu} \pi_{\mu\nu} }{2 T^2 ({\cal E}{+}{\cal P})} f_0 (1{+}f_0).
	\label{cooper_frye_defn2}
	\end{eqnarray}
	Here the distribution function on $\Sigma_\mathrm{f}$ is written as the sum of a local equilibrium distribution $f_0$ and a viscous correction $\delta f$. $f_0$ is a function of the (space-time dependent) temperature $T$, chemical potential $\mu$, and hydrodynamic flow velocity $u^\mu(x)$ on $\Sigma_\mathrm{f}$. The viscous correction $\delta f$ \cite{Teaney:2003kp,Dusling:2009df} depends on the shear stress tensor $\pi^{\mu\nu}$ on $\Sigma$, normalized by the enthalpy density $\mathcal{E}{+}\mathcal{P}$ (where $\mathcal{E}(T,\mu)$ and $\mathcal{P}(T,\mu)$ are the energy density and thermal pressure of the fluid, respectively) and contracted with the particle momentum $p$, as well as on a scalar function $\chi(p^2)$ whose specific form varies with the properties of the interactions among the constituents at freeze-out and which we take as $\chi(p^2)\equiv1$ \cite{Dusling:2009df}.  
	
	If the two identical particles are emitted independently, their quantum-statistical correlations can be expressed through this single-particle emission function as 
	\begin{equation}
		C(\vec{q}, \vec{K}) \approx  1 + \l| \frac{ \int d^4 x \, S(x,K) e^{iq \cdot x} }{\int d^4 x \, S(x,K)} \r|^2.
		\label{corrfunc_vs_S_defn}
	\end{equation} 
	The nature of the approximations indicated by the ``$\approx$'' sign is explained in the many available reviews of the formalism \cite{Heinz:1996bs,Heinz:1999rw,Wiedemann:1999qn,Lisa:2005dd,Lisa:2008gf}; for high-energy collisions between heavy ions they are quite accurate.
	
For a spatially Gaussian emission function the $q$-dependence of $C(\vec{q}, \vec{K})$ is Gaussian, and the (inverse) width parameters of this Gaussian (i.e. the HBT radii) can be directly extracted from its curvature at the origin $q=0$, giving the following relations \cite{Chapman:1994ax} between the HBT radii and the space-time variances of the source function $S(x,K)$:
	\begin{equation}
		R^2_{ij} = \Savg{(x_i - \beta_i t)(x_j - \beta_j t)} - \Savg{x_i - \beta_i t}\Savg{x_j - \beta_j t}.
			\label{Chap2_svHBT_defn}
	\end{equation}
	Here the average $\Savg{\dots}$ denotes the space-time average over the emission function,
	\begin{equation}
		\Savg{f(x)} \equiv \frac{\int d^4 x\, f(x)\, S(x,K)}{\int d^4 x\, S(x,K)},
			\label{Chap2_sv_avg_defn}
	\end{equation}
	which depends on the pair momentum $K$. This method for computing the HBT radii is called the source-variances (SV) method. 
	
	Both the SV method and the Gaussian Fit (GF) method described in the preceding subsection become unreliable if the $q$-dependence of the correlation function $C(\vec{q}, \vec{K})$ exhibits strong deviations from a simple Gaussian shape. As we discuss in detail below, the latter is the case when decay products from a large set of resonances with a range of lifetimes contribute to the particles used for the HBT measurement.

	%%%%%%%%%%%%%%%%%%%%%%%%%%%%%%%%%%%%%%%%%%%%%%%%%%%%%%%
	\subsection{Resonance decay contributions}
	\label{Sec2d}
	%%%%%%%%%%%%%%%%%%%%%%%%%%%%%%%%%%%%%%%%%%%%%%%%%%%%%%%
	For sudden freeze-out on a surface $\Sigma$, particles (say, pions) are produced in two ways:
	\begin{enumerate}
		\item direct thermal emission from $\Sigma$ according to Eq.~(\ref{cooper_frye_defn1}),
		\item indirect emission through decays $r\to\pi$ of unstable resonances which themselves are
		         emitted according to Eq.~(\ref{cooper_frye_defn1}).
	\end{enumerate}
	The full emission function is given as the sum of all these contributions:\footnote{
	Note that the sum over resonances $r$ includes sums over all intermediate resonance
	states which can decay to, say, $\pi^+$s: e.g., $r \to r_1 \to \pi^+$, $r \to r_2 \to r_3 \to \pi^+$,
	and so on. Below, $S_{r\rightarrow \pi^+}$ is calculated explicitly for the case when the resonance $r$
	is produced thermally.  The generalization to resonances with other phase-space distributions
	is straightforward, by substituting the latter for $S_r^\mathrm{dir}(x,p)$ under the integral over
	the decay phase-space. The interested reader will find a more thorough discussion of our procedure in
	Ref.~\cite{Qiu:2012tm}.
	}
	\begin{equation}
		S(x,p) \equiv S_\pi(x,p) = S^{\mathrm{dir}}_{\pi}(x,p) + \sum_r S_{r \rightarrow \pi}(x,p). 
	\label{resonance_decay_emission_fcn}
	\end{equation}

	The efficient calculation of the resonance decay contributions to the two-pion correlation function, and the study of their effect on its shape, are the main goals of this work. We therefore briefly outline the calculation of $S_{r \rightarrow \pi}(x,p)$ in the rest of this subsection, following the notation of Ref.~\cite{Wiedemann:1996ig}. The components of the momentum $P$ of the parent resonance is denoted by capital letters while those of the momentum $p$ of the daughter pion are labeled by lower-case letters.\footnote{
	All particle information for the resonances included in our calculation is derived from the Particle Data Group (PDG) Review of Particle Physics \cite{Agashe:2014kda}.  We have tabulated the necessary particle data (e.g., branching ratios) in a form which was originally developed as a part of the Azhydro calculation \cite{azhydroPaper}.  This tabulated data is included as part of the Github account where the entire code used in the present work is available for download \cite{myPDGfile}.
	}
	
	We work in the $o,s,l$ coordinate system defined by the momentum $\vec{p}$ of the daughter pion. For the calculation of the two-particle correlation function we need $S_\pi(x,p)$ at the pair momentum $\vec{K}$, $p \mapsto K$. In order to avoid confusion between capital and lower-case letters (the capital letter $K$ is not associated with the resonance, but with its decay products), we will make this substitution only at the very end.

	In this coordinate system the four-vectors describing the parent resonance and daughter pion are given by	
	 \begin{eqnarray}
	 \label{eq:P}
	         P^\mu &\equiv& (E_{_P}, P_o, P_s, P_{l}) \\
			&=&
   	         \bigl( M_\perp \cosh Y,\, P_\perp \cos\Phi,\,P_\perp \sin\Phi,\, M_\perp \sinh Y \bigr) ,
         \nonumber\\
         \label{eq:p}
	        p^\mu &\equiv& (E,\, p_o,\, p_s,\, p_{l}) \\
			& \equiv & 
	        \bigl( m_\perp \cosh y,\, p_\perp, \, 0,\, m_\perp \sinh y \bigr) . \nonumber
	 \end{eqnarray}
	The angle $\Phi$ in Eq.~(\ref{eq:P}) is the azimuthal angle between the transverse momenta of the parent resonance and daughter pion. 
	
	In the parent resonance rest frame (with variables denoted by a $^*$), the daughter particle has energy and momentum 
	 \begin{eqnarray}
	   E^* &=& \sqrt{m^2 + {p^*}^2}\, ,
	   \nonumber \\
	   p^* &=& \frac{\sqrt{[(M+m)^2-s][(M-m)^2-s]} }{ 2M}\, ,
	 \end{eqnarray}
	where $s = \left(\sum_{i=2}^n p_i \right)^2$ is the squared invariant mass of the remaining $n{-}1$ daughter particles produced in the (assumed) $n$-body decay. It can vary between $s_- = \left( \sum_{i=2}^n m_i \right)^2$ and $s_+ = (M-m)^2$. $g(s)$ is the decay phase space for the $(n{-}1)$ unobserved particles. For two- and three-body decays, the latter reads
	\begin{itemize}
	\item
	for two-body decays:
	  \begin{eqnarray}
	    g(s) &=& \frac{b }{ 4\pi p^*} \delta \left( s - m_2^2 \right)\, ,
	  \end{eqnarray}
	 
	\item
	for three-body decays ($s_- = (m_2+m_3)^2,\, s_+ = (M-m)^2$):
	  \begin{equation}
	        g(s) = \frac{M b }{ 2\pi s} 
	        \frac{\sqrt{[s - (m_2{+}m_3)^2][s - (m_2{-}m_3)^2]}}
	        { Q(M,m,m_2,m_3)}\, .
	  \end{equation}
	\end{itemize}
	To obtain the daughter pion emission function at momentum $p$ we must integrate the decay phase space over all contributing momenta $P$ of the parent resonance. We introduce integration variables $v\in [-1,1]$, $\zeta \in [-\pi,\pi]$ by writing
	  \begin{eqnarray}
	    M_\perp &=& \overline{M}_\perp + \Delta M_\perp \, \cos\zeta\, ,
	  \\
	    Y &=& y  + v\, \Delta Y \, ,
	  \end{eqnarray}
	where $\Delta M_\perp$ and $\Delta Y$ are obtained from the kinematic limits associated with the decay through the following relations \cite{Wiedemann:1996ig}
	  \begin{eqnarray}
	        M_{\perp,\pm} &=& \overline{M}_\perp \pm \Delta M_\perp
	        \equiv \frac{ E^* M m_\perp \cosh(Y{-}y) }{
	            m_\perp^2 \cosh^2(Y{-}y) - p_\perp^2 } 
	          \nonumber \\
	         && \pm 
	          \frac{M p_\perp 
	           \sqrt{ {E^*}^2 + p_\perp^2 - m_\perp^2 \cosh^2(Y{-}y)}
	           }{
	           m_\perp^2 \cosh^2(Y{-}y) - p_\perp^2 } ,
	 \\ 
	        Y_\pm &=& y \pm \Delta Y \equiv 
	        y \pm \ln \left( \frac{p^* }{ m_\perp} 
	                       + \sqrt{ 1 + \frac{{p^*}^2 }{ m_\perp^2} } 
	                  \right) .
	  \end{eqnarray}
	With these definitions the contribution to Eq.~\eqref{resonance_decay_emission_fcn} from the decay of resonance $r$ can be written as \cite{Wiedemann:1996ig}
	\begin{equation}
		S_{r\rightarrow \pi}(x,p) = \sum_{k=\pm} \int_{\mathbf{R}} 
		\int^{\infty}_0 \Gamma\,d\tau\, e^{-\Gamma \tau} \,S^{\dir}_{r}\l( x-\frac{P^k \tau}{M}, P^k \r), 
	\label{res_decay_cont}
	\end{equation}
	where $\Gamma$ is the width of the resonance $r$ and
	\begin{eqnarray}
		\int_{\mathbf{R}}
			& \equiv & M \int^{s_+}_{s_-} ds\,g(s) 
		\int^{+1}_{-1}\frac{\Delta Y dv}{\sqrt{m_{\perp}^2 \cosh^2(v \Delta Y) - p_{\perp}^2}} \nonumber\\
			& \times & \int^{\pi}_0 
					d\zeta \, \l( \overline{M}_{\perp} + \Delta M_{\perp} \cos \zeta \r) .
	\label{phase_space_integrals}
	\end{eqnarray}
	The sum over $k=\pm$ in Eq.~\eqref{res_decay_cont} corresponds to the following two solutions of the energy-momentum constraints between the parent and daughter momenta \cite{Wiedemann:1996ig}:
	\begin{equation}
	P^{\pm}{\,\equiv\,}\left( M_\perp \cosh Y, P_\perp \cos \Phi_{\pm}, 
	                       P_\perp \sin \Phi_{\pm}, M_\perp \sinh Y \right),\quad
	\label{Ppm_defn}
	\end{equation}
	where
	\begin{equation}
	\Phi_{\pm} \equiv \pm \tilde{\Phi} \ \ \text{with} \ \cos \tilde{\Phi} \equiv \frac{m_{\perp} M_{\perp} \cosh(Y{-}y) - E^* M}{p_{\perp} P_{\perp}}.
	\label{Phipm_defn}
	\end{equation}
	%

	%%%%%%%%%%%%%%%%%%%%%%%%%%%%%%%%%%%%%%%%%%%%%%%%%%%%%%%
	\subsection{Resonance decay effects on the HBT radii from the SV method}
	\label{Sec2d1}
	%%%%%%%%%%%%%%%%%%%%%%%%%%%%%%%%%%%%%%%%%%%%%%%%%%%%%%%
	
	In this and the following subsection we describe the effects of resonance decays on the HBT radii when using the source-variances (SV) and Gaussian fitting (GF) methods for their computation.	
	
	In the SV method, Eqs.~\eqref{Chap2_svHBT_defn} and \eqref{Chap2_sv_avg_defn} show that computing the HBT radii involves evaluating the following integrals:\iffalse\footnote{%
		In this and the following subsection, we return to the standard Cartesian coordinate 
		system with $x \equiv \l(x^0,x^1,x^2,x^3\r) \equiv \l( t, \mathrm{x, y,} z \r)$. We use 
		roman letters for the coordinates x and y, to distinguish the coordinate y from the 
		momentum rapidity $y$. In this coordinate system, the transverse momentum of the 
		daughter pion has momentum $\vec{p}_\perp=(p_x,p_y)=p_\perp(\cos\phi_p,\sin\phi_p)$.		
		We also introduce the shorthand notation $\int_x \dots \equiv \int d^4x \dots$.}  \fi
	%
	\begin{eqnarray}
		\int_x \, S(x,p), \ 
		\int_x \, x^{\mu} S(x,p),\ 
		\int_x \, x^{\mu}x^{\nu} S(x,p).
		\label{general_SV_form}
	\end{eqnarray}
	In this and the following subsection, we return to the standard Cartesian coordinate 
		system with $x \equiv \l(x^0,x^1,x^2,x^3\r) \equiv \l( t, \mathrm{x, y,} z \r)$. We use 
		roman letters for the coordinates x and y, to distinguish the coordinate y from the 
		momentum rapidity $y$. In this coordinate system, the transverse momentum of the 
		daughter pion has momentum $\vec{p}_\perp=(p_x,p_y)=p_\perp(\cos\phi_p,\sin\phi_p)$.		
		We also introduce the shorthand notation $\int_x \dots \equiv \int d^4x \dots$.
	
	 Substituting Eq.~\eqref{resonance_decay_emission_fcn} for the full emission function $S(x,p)$ 
	 into, say, the last expression one finds 
	\begin{eqnarray}
		\int_x \, x^{\mu}x^{\nu} S(x,p)
			&=& \int_x \, x^{\mu}x^{\nu} S^{\dir}_{\pi}(x,p) \nonumber\\
			&& + \sum_r \int_x \, x^{\mu}x^{\nu} S_{r \to \pi}(x,p),
		\label{subbed_SV_form}
	\end{eqnarray}
	with similar expressions for the other integrals above. The direct contribution in the first term is a standard Cooper-Frye integral and straightforwardly evaluated with existing tools. We now show how to simplify the sum over resonance contributions in the second term. 
	
	\begin{widetext}
	Substituting Eqs.~\eqref{cooper_frye_defn1}, \eqref{res_decay_cont} and \eqref{phase_space_integrals} into Eq.~\eqref{subbed_SV_form} and using the integration over $x$ to eliminate the $\delta$-function in \eqref{cooper_frye_defn1} we find 
	\begin{eqnarray}
		%\int_x x^{\mu}x^{\nu}  S_{r\rightarrow \pi}(x,p) & = & \frac{1}{\l( 2\pi \r)^3} \sum_{k=\pm}
		%	\int_{\mathbf{R}}
		%	\int^{\infty}_0 \Gamma\, d\tau\, e^{-\Gamma \tau} \int_{\Sigma} P^k \cdot d^3 \sigma(\tilde{x}) \\
		%& & \times \l(\tilde{x}^{\mu} + \l( \frac{P^k}{M} \r)^{\mu} \tau\r)\l(\tilde{x}^{\nu} + \l( \frac{P^k}{M} \r)^{\nu} \tau\r) f_r(\tilde{x}, P^k). \nonumber
		\!\!\int_x x^{\mu}x^{\nu}  S_{r\rightarrow \pi}(x,p) & = & \frac{1}{\l( 2\pi \r)^3} \sum_{k=\pm}
			\int_{\mathbf{R}}
			\int^{\infty}_0 \!\Gamma d\tau\, e^{-\Gamma \tau} \!\int_{\Sigma} \! P^k {\cdot} d^3 \sigma(\tilde{x})
			\l(\tilde{x}^{\mu} {+} \l( \frac{P^k}{M} \r)^{\!\mu} \!\tau\r)\l(\tilde{x}^{\nu} {+} \l( \frac{P^k}{M} \r)^{\!\nu} \!\tau\r) f_r(\tilde{x}, P^k).
	\label{long_equation_subbed}
	\end{eqnarray}
	The $\tau$-integral can be done analytically leading to 
	\begin{eqnarray}
	\!\!\!\!\!
		\int_x \, x^{\mu}x^{\nu} S_{r\rightarrow \pi}(x,p) & = & \sum_{k=\pm}
			\int_{\mathbf{R}}
			\l[ \l\lbrace \tilde{x}^{\mu} \tilde{x}^{\nu} \r\rbrace ^k_r  + \alpha^{\mu}_k \l\lbrace 
			\tilde{x}^{\nu}\r\rbrace ^k_r + \alpha^{\nu}_k \l\lbrace \tilde{x}^{\mu}\r\rbrace ^k_r + 
			2\alpha^{\mu}_k \alpha^{\nu}_k \l\lbrace 1 \r\rbrace ^k_r \r], \quad
		\label{ymuynuintegrals}
	\end{eqnarray}
	where we introduced the shorthands
	\begin{eqnarray}
	\alpha^{\mu}_k
		& \equiv & \l( P^k \r)^{\mu}/(\Gamma M), 
		\label{alpha_defn}\\
	\l\lbrace \dots \r\rbrace ^k_r
		& \equiv & \frac{1}{\l( 2\pi \r)^3} \int_{\Sigma} P^k{\,\cdot\,}d^3\sigma(\tilde{x}) \,\{\dots\}\, 
		f_r(\tilde{x}, P^k) = \int_x \{\dots\}\, S^{\dir}_r(x,P^k). 
		\label{stm_1_defn}
	\end{eqnarray}
	\end{widetext}
	Similarly
	\begin{eqnarray}
		&&\int_x x^{\mu} S_{r\rightarrow \pi}(x,p) = \sum_{k=\pm}
			\int_{\mathbf{R}}
			\l[ \l\lbrace \tilde{x}^{\mu} \r\rbrace ^k_r  + \alpha^{\mu}_k \l\lbrace 1 \r\rbrace ^k_r \r], 
		\label{ymuintegrals}
	\\	
		&&\int_x S_{r\rightarrow \pi}(x,p) = \sum_{k=\pm}
			\int_{\mathbf{R}}
			 \l\lbrace 1 \r\rbrace ^k_r.
		\label{unityintegrals}
	\end{eqnarray}  
	We refer to $\l\lbrace 1 \r\rbrace ^k_r$, $\l\lbrace \tilde{x}^{\mu} \r\rbrace ^k_r$, and $\l\lbrace \tilde{x}^{\mu} \tilde{x}^{\nu} \r\rbrace ^k_r$ as \textit{space-time moments} of the single-particle distribution for the resonance $r$, evaluated at momentum $P^k$.  In general, each space-time moment possesses a three-dimensional dependence on the momentum $P^k$, including the two-dimensional transverse momentum $\vec{P}^k_{\perp}$ and the rapidity $Y$ characterizing the longitudinal motion.  The additional assumption of boost-invariance, however, allows us to simplify the problem somewhat further, by enabling us to separate out the dependence on $Y$.
	
	Let us illustrate this with a few examples. Boost invariance requires that all spatial observables (e.g., $T^{\mu\nu}(x)$) be independent of the space-time rapidity
	\begin{equation}
		\eta_s \equiv \frac{1}{2} \ln \l( \frac{t+z}{t-z} \r)
	\end{equation} 
	and that all momentum-space observables (e.g., $E_p \l({dN / d^3p}\r)$) are independent of 
	the longitudinal momentum-space rapidity
	\begin{equation}
		y \equiv \frac{1}{2} \ln \l( \frac{E_p + p_z}{E_p - p_z} \r).
	\end{equation}  
	Distribution functions such as $f(x,p)$ and $S(x,p)$ are allowed to depend only on the 
	difference $\eta_s{-}y$. Then
	\begin{eqnarray}
		E_p\frac{d N}{d^3p} & = & \int_x S(x,p)
		\label{single_particle_spectra}
		\nonumber\\
		&=& \int d^2r_\perp \int^{\infty}_0 \tau d\tau \int^{\infty}_{-\infty} d\eta_s \,
		S(\vec{r}_{\perp},\tau,\vec{p}_{\perp},\eta_s{-}y) 
		\nonumber\\
		& = & \int d^2r_\perp \int^{\infty}_0 \tau d\tau \int^{\infty}_{-\infty} d\tilde{\eta}_s \,
		S(\vec{r}_{\perp},\tau,\vec{p}_{\perp},\tilde{\eta}_s)
	\end{eqnarray} 
	is automatically $y$-independent.
	
	For source variances that depend on space-time rapidity, however, eliminating the $y$-independence by a shifting the space-time rapidity under the integral is no longer possible. For instance 
	\begin{widetext}
	\begin{eqnarray}
		\avg{t} \equiv \avg{\tau\cosh\eta_s}\l( \vec{p}_{\perp},y \r)
		& \equiv & \int d^2r_\perp \int^{\infty}_0 
		\tau^2 d\tau \int^{\infty}_{-\infty} d\eta_s \,\cosh\eta_s \, 
		S(\vec{r}_{\perp},\tau,\vec{p}_{\perp},\eta_s{-}y) /\{1\}
		\nonumber\\
		&=& \int d^2r_\perp \int^{\infty}_0 \tau^2 d\tau \int^{\infty}_{-\infty} d\tilde{\eta}_s \,
		\cosh\l(\tilde{\eta}_s+y\r) \, S(\vec{r}_{\perp},\tau,\vec{p}_{\perp},\tilde{\eta}_s) /\{1\}
		\nonumber\\
		&=& \cosh y\, \avg{t}_{y=0} + \sinh y\,  \avg{z}_{y=0},
		\label{t_vs_y_p}
	\end{eqnarray}  
where $\{1\}\equiv\{1\}(p)\equiv \int_x S(x,p)$ is $y$-independent. Similar identities hold for the other source variances which depend on $t$ or $z$. Explicitly, they are: 
	\begin{eqnarray}
		\avg{z}
			&\equiv & \cosh y\,  \avg{z}_{y=0} + \sinh y\,  \avg{t}_{y=0}, 
			\label{z_vs_y_p}\\
		\avg{t^2}
			&\equiv & \cosh^2 y\,  \avg{t^2}_{y=0} + \sinh (2y) \, \avg{z t}_{y=0} +
			   \sinh^2 y\,  \avg{z^2}_{y=0}, 
			   \label{t2_vs_y_p}\\
		\avg{z^2}
			&\equiv & \cosh^2 y \, \avg{z^2}_{y=0} + \sinh(2y)\,  \avg{z t}_{y=0} + 
			   \sinh^2 y\,  \avg{t^2}_{y=0},
			   \label{z2_vs_y_p}\\
		\avg{z t}
			&\equiv & \cosh(2y)\,  \avg{z t}_{y=0} + 
			  \sinh(2y) \,\frac{1}{2} \Bigl(\avg{z^2} + \avg{t^2} \Bigr)_{y=0},
			  \label{zt_vs_y_p}\\
		\avg{\mathrm{y} z}
			&\equiv & \cosh y\,  \avg{\mathrm{y} z}_{y=0} + \sinh y \, \avg{\mathrm{y} t}_{y=0},
			\label{yz_vs_y_p}\\
		\avg{\mathrm{x} z}
			&\equiv & \cosh y\,  \avg{\mathrm{x} z}_{y=0} + \sinh y \, \avg{\mathrm{x} t}_{y=0},
			 \label{xz_vs_y_p}\\
		\avg{\mathrm{y} t}
			&\equiv & \cosh y \, \avg{\mathrm{y} t}_{y=0} + \sinh y\,  \avg{\mathrm{y} z}_{y_p=0},
			 \label{yt_vs_y_p}\\
		\avg{\mathrm{x} t}
			&\equiv & \cosh y \, \avg{\mathrm{x} t}_{y=0} + \sinh y \, \avg{\mathrm{x} z}_{y=0}.
			\label{xt_vs_y_p}
	\end{eqnarray}  
	\end{widetext}
	The remaining source variances do not depend on either $t$ or $z$ and are thus (according to Eq.~(\ref{single_particle_spectra})) independent of $y$. These equations show that the $y$-dependence of the source variances can be calculated trivially from their values at midrapitity $y\eq0$. Hence, for boost-invariant sources, we need to compute the space-time moments of the source only on a \textit{two}-dimensional grid at $y=0$.
	
	The source variances including resonance decay contributions can now be written as 
	\begin{widetext}
	\begin{eqnarray}
	 	\label{eq:vars}
		\langle x^\mu x^\nu \rangle(p)
		&=&
		\int_x\, x^\mu x^\nu S(x,p)/\{1\}(p)
		\\\nonumber
	 	&=& \left[\int_x\, x^\mu x^\nu S^{\mathrm{dir}}_{\pi}(x,p) 
			+ \sum_r \sum_{k=\pm} \int_{\mathbf{R}} \int_x 
			\bigl( x^\mu x^\nu{+}\alpha^\mu_k x^\nu{+}\alpha^\nu_k x^\mu{+}2 \alpha^\mu_k 
			    \alpha^\nu_k \bigr)\, S^{\dir}_r\bigl(x,P^k\bigr) \right]\Big/\{1\}(p)
	\end{eqnarray}
	where $\alpha^\mu_k$ is defined in Eq.~(\ref{alpha_defn}) and
	\begin{equation}
	\{1\}(p) = \int_x S^{\mathrm{dir}}_{\pi}(x,p) 
	            + \sum_r \sum_{k=\pm} \int_{\mathbf{R}} \int_x S^{\dir}_r(x,P^k) 
	            = \{1\}^\mathrm{dir} +  \sum_r \sum_{k=\pm} \int_{\mathbf{R}} \{1\}_r^k.
	\end{equation}
	\end{widetext}
	Equation~(\ref{eq:vars}) exposes most clearly the optimal way of structuring the calculation of the source variances with resonances, since each term under the sum of decay phase-space integrals over resonance emission functions has the generic form given in Eq.~\eqref{stm_1_defn}.  We outline this approach in Algorithm \ref{SVHBT_algorithm}.
	\begin{widetext}
	\vspace*{5mm}
	\begin{algorithm}[!htp]
	\caption{Efficiently compute source variances with resonance decay contributions}
		\begin{algorithmic}[1]
		\ForAll{resonance $r$ and $\pi^+$}
		\State Compute the set of quantities $\int_x \,S^{\dir}_r(x,p),$ $\int_x \, x^{\mu} \, S^{\dir}_r(x,p),$ $\int_x \, x^{\mu} x^{\nu} \, S^{\dir}_r(x,p)$ on a two-dimensional grid in $\vec{p}_{\perp}$ (e.g., $p_{\perp}$ and $\phi_p$), and use Eqs.~\eqref{t_vs_y_p}-\eqref{xt_vs_y_p} to obtain the dependence on the rapidity $y$
		\State Use these grids to evaluate the various terms in the quantities \eqref{general_SV_form} as described in Eq.~\eqref{eq:vars}
		\EndFor
		\State Sum the thermal and resonance decay contributions to obtain the full set of quantities $\int_x S(x,p),$ $\int_x x^{\mu} \, S(x,p),$ and $\int_x x^{\mu} x^{\nu} \, S(x,p)$
		\State Use these quantities to construct the full source variances, e.g., \[ \Savg{x z} = \frac{\int_x x \, z \, S(x,p)}{\int_x S(x,p)} \]
			\State Compute the HBT radii from the complete set of the full source variances 
		\end{algorithmic}
		\label{SVHBT_algorithm}
	\end{algorithm}
	%\end{widetext}
	%
	Since the source variances are here computed in the laboratory-fixed Cartesian coordinate system $(t,\mathrm{x,y,}z)$ while the HBT radii are defined and measured in the $osl$-coordinate system, the last step in Algorithm \ref{SVHBT_algorithm} contains an implicit transformation between these two coordinate systems. For this step we substitute for $\vec{p}$\, the pair momentum $\vec{K}=(K_\perp\cos\Phi_K,K_\perp\sin\Phi_K,K_l)$. The necessary transformation rules are then (see Fig.~\ref{fig:oslcoords_with_r})
	\begin{eqnarray}
		x_o &=& r \cos(\phi{-}\Phi_K) = \mathrm{x} \cos \Phi_K + \mathrm{y} \sin \Phi_K, \\
		x_s &=& r \sin(\phi{-}\Phi_K) = - \mathrm{x} \sin \Phi_K + \mathrm{y} \cos \Phi_K, \\
		x_l &=& z.
		\label{osl_xyz_transf}
	\end{eqnarray}
	Thus we find, for instance, 
	%
	%\begin{widetext}
	\begin{eqnarray}
		\Savg{x_o x_l} &=& \cos\Phi_K\langle\mathrm{x} z\rangle + \sin\Phi_K\langle\mathrm{y}z\rangle
		\nonumber\\
				        &=& \cos\Phi_K \frac{\int_x \mathrm{x} z\, S(x;K_T,\Phi_K)}
		                                                    {\int_x S(x;K_T,\Phi_K)} 
					+ \sin\Phi_K \frac{\int_x \mathrm{y} z \,S(x;K_T,\Phi_K)}
					                           {\int_x S(x;K_T,\Phi_K)},
	\end{eqnarray}
	\end{widetext}
	and similarly for the other source variances in the $osl$-coordinates.  The HBT radii are then determined by inserting this set of quantities into the expression \eqref{Chap2_svHBT_defn}.

	%%%%%%%%%%%%%%%%%%%%%%%%%%%%%%%%%%%%%%%%%%%%%%%%%%%%%%%
	\subsection{Resonance decay effects on the HBT radii from the GF method}
	\label{Sec2d2}
	%%%%%%%%%%%%%%%%%%%%%%%%%%%%%%%%%%%%%%%%%%%%%%%%%%%%%%%
	
	To apply the Gaussian Fit method we must first compute the correlation function \eqref{corrfunc_vs_S_defn}, by Fourier transforming the full emission function (\ref{resonance_decay_emission_fcn}).  Starting from Eq.~\eqref{res_decay_cont} we find
	  \begin{eqnarray}
	     \tilde S_{r\to\pi}(q,p) 
	     &=& \int_x e^{iq{\cdot}x}\, S_{r\to\pi}(x,p)
	     \nonumber \\
	        &=& \sum_\pm \int_{\bf R} \int_0^{\infty} d(\Gamma\tau)\, 
	        \exp\left[ -\Gamma\tau \left( 1 - i\frac{q{\cdot}P^\pm }{ M \Gamma}
	                               \right) \right] \nonumber\\
	     	&& \times \int_x e^{iq{\cdot}x}\, S_r^{\rm dir}(x,P^\pm)
	     \nonumber \\
	        &=& \sum_\pm \int_{\bf R}
	        \frac{1 }{
	         1 - i \frac{ q{\cdot}P^\pm }{ M\Gamma}}\, \tilde S_r^{\rm dir}(q,P^\pm) \, ,
	        \label{FTspectra}
	  \end{eqnarray}
	where in the first step we shifted the $x$-integration variable before performing the $\tau$-integration.

	 As noted in Eq.~(\ref{On_shell_condition_for_pions}), the four components of $q$ are not independent, but constrained by orthogonality to the pair momentum $K$:
	\begin{equation}
		q^0 = \vec{\beta} \cdot \vec{q}, \quad \vec{\beta} = \vec{K}/K^0 
		\approx \vec{K}/E_K = \frac{\vec{K}}	{\sqrt{m^2_\pi + \vec{K}^2}}. 
		\label{qtConstraint}
	\end{equation}
	Writing 
	\begin{equation}
		\vec{p}_1 = \vec{K} + \frac{\vec{q}}{2},\,\, \vec{p}_2 = \vec{K} - \frac{\vec{q}}{2},
	\end{equation}
	we obtain the useful relation
	\begin{eqnarray}
		q^0 & \equiv & E_1 - E_2 
			  = \sqrt{m_\pi^2 + \vec{p}_1^2} - \sqrt{m_\pi^2 + \vec{p}_2^2} \nonumber\\
		       & = & \sqrt{m_\pi^2 + \vec{K}^2 + \frac{1}{4} \vec{q}^2 + \vec{q}\cdot \vec{K}} \nonumber\\
			&& - \sqrt{m_\pi^2 + \vec{K}^2 + \frac{1}{4} \vec{q}^2 - \vec{q}\cdot \vec{K}}. \\
		\label{qt_useful_reln}
	\end{eqnarray}
	The Fourier transform is therefore not fully four-dimensional since $q^0$ is not an independent degree of freedom.\footnote{
	This is the underlying reason why a three-dimensional set of HBT radii in the
	$osl$-coordinate system requires a set of source variances characterizing the source
	function in a four-dimensional Cartesian coordinate system: since the Fourier transform
	is only three-dimensional, thanks to the constraint on $q^0$, it can only relay
	three-dimensional information regarding the source structure.  This means that the
	$R^2_{ij}$, in general, necessarily represent non-trivial convolutions of the spatial
	\emph{and} temporal structure of the freeze-out surface \cite{Heinz:1999rw}, so that an 
	exclusively \emph{geometric} interpretation of the HBT radii will almost always
	produce insights which are either misleading or simply incorrect.
	}  
	%\begin{widetext}
	Using Eq.~\eqref{qtConstraint} to eliminate $q^0$ from Eq.~(\ref{FTspectra}), the correlation function \eqref{corrfunc_vs_S_defn} can be written in terms of the on-shell Fourier transform of the emission function as
	%\begin{equation}
	%	C(\vec{q},\vec{K})
	%		= 1 +
	%		\frac{\l| \tilde{S}^{\mathrm{dir}}_{\pi}(\vec{q},\vec{K}) \r|^2
	%			+ 2 \sum_r \re \l[\tilde{S}^{\mathrm{dir}}_{\pi}(\vec{q},\vec{K}) 
	%			       \tilde{S}^*_{r \to \pi}(\vec{q},\vec{K}) \r]
	%			+ \l| \sum_r \tilde{S}_{r \to \pi}(\vec{q},\vec{K}) \r|^2}
	%			{\l| \tilde{S}^{\mathrm{dir}}_{\pi}(0,\vec{K}) \r|^2
	%			+ 2 \sum_r \re \l[\tilde{S}^{\mathrm{dir}}_{\pi}(0,\vec{K}) 
	%			       \tilde{S}^*_{r \to \pi}(0,\vec{K}) \r]
	%			+ \l| \sum_r \tilde{S}_{r \to \pi}(0,\vec{K}) \r|^2}. 
	%\label{CFWR_defn}
	%\end{equation}
	\begin{equation}
		C(\vec{q},\vec{K}) = 1 + \frac{N(\vec{q},\vec{K})}{N(0,\vec{K})},
	\label{CFWR_defn}
	\end{equation}
	where\footnote{
	We note that this expression corrects a typographical error in \cite{Wiedemann:1996ig}
	which omitted the complex conjugation from the last term in Eq.~(\ref{eq57}).
	}
	%
	%\begin{widetext}
	\begin{eqnarray}
	\label{eq57}
		N(\vec{q},\vec{K})
			&\equiv& \l| \tilde{S}^{\mathrm{dir}}_{\pi}(\vec{q},\vec{K}) \r|^2 + \l| \sum_r \tilde{S}_{r \to \pi}(\vec{q},\vec{K}) \r|^2 \nonumber\\
			&& + 2 \sum_r \re \l[\tilde{S}^{\mathrm{dir}}_{\pi}(\vec{q},\vec{K}) 
				       \tilde{S}^*_{r \to \pi}(\vec{q},\vec{K}) \r].\quad
	\end{eqnarray}
	%\end{widetext}
	%
	After Eq.~\eqref{CFWR_defn} has been computed the GF HBT radii are obtained by fitting to the functional form \eqref{corrfunc_functional_defn}. To compute the thermal pion GF HBT radii one keeps only the first term in the numerator and denominator of Eq.~(\ref{CFWR_defn}).
	
The on-shell constraint on $q^0$ entails a subtlety for the numerical evaluation of the decay phase-space integrals in Eq.~(\ref{FTspectra}) that requires discussion. As described in the Appendix, these integrals are computed by interpolating a precomputed momentum-space array of Fourier-transformed emission functions $\tilde S_r^{\rm dir}(q,P^\pm)$. If we use the on-shell constraint for $q^0$ before computing this array, it will be 8-dimensional, labeled by $q_x,q_y,q_z;P^k_T,\Phi_P,Y_P$ as well as additionally by $K_T$ and $\Phi_K$ through the constraint (\ref{qtConstraint})\footnote{Although our code can compute the HBT radii at any longitudinal pair momentum $K_L$, for simplicity we consider only mid-rapidity pions $\l(  K_L = 0\r)$ in this paper.}. It is more economical to instead leave $q^0$ initially unconstrained and evaluate $\tilde S_r^{\rm dir}(q,P^\pm)$ on a 7-dimensional grid $(q_x,q_y,q_z,q^0; P^k_T,\Phi_P, Y_P)$, interpolating $q^0$ to the desired value $q^0=\vec{q}{\cdot}\vec{K}/E_K$ only at the end of the calculation. Details of the algorithm for computing Eq.~(\ref{CFWR_defn}) are found in the Appendix.

	%%%%%%%%%%%%%%%%%%%%%%%%%%%%%%%%%%%%%%%%%%%%%%%%%%%%%%%
	\subsection{Gaussian Fit procedure}
	\label{Sec2e}
	%%%%%%%%%%%%%%%%%%%%%%%%%%%%%%%%%%%%%%%%%%%%%%%%%%%%%%%
	
	By definition, the GF HBT radii must be determine by fitting the correlation function \eqref{corrfunc_vs_S_defn} to the functional form \eqref{corrfunc_functional_defn}.  The challenges of performing such fits are already well-documented \cite{Frodermann:2006sp}, and should be matched as closely as possible to the experimental procedure.  This procedure comprises several key ingredients, including the following:
	\begin{enumerate}
		\item \textit{Distinguishing between one-dimensional and three-dimensional Gaussian fits.}  In principle, there are different ways of fitting of the correlation function.  One-dimensional fits are performed along a slice of the correlation function along some axis in $q$-space.  By construction, such a fit optimally represents just this slice in a Gaussian form, without constraints from other directions in $q$-space.  By contrast, a three-dimensional fit must represent not just the correlation function slices along each axis in $q$-space, but must also fit as closely as possible points which lie off-axis as well.  One-dimensional and three-dimensional fits therefore yield somewhat different results, and it is crucial to recognize that only the latter correspond to the most general, three-dimensional HBT analyses which experimentalists perform \cite{Adams:2004yc}.
		It is easy to appreciate how these differences originate.  Consider a correlation function evaluated in $N_q^3$ bins in $q$-space, with $N_q$ bins along each axis, each spaced from $-q_{\mathrm{max}}$ to $+q_{\mathrm{max}}$, for simplicity.  Now, consider separating the $q$-bins into those with $|q| \geq q_{\mathrm{max}}/2$ and $|q|<q_{\mathrm{max}}/2$ along each $q$-axis.  Clearly, for the one-dimensional fits along each axis, assuming the $q$-bins are equally spaced and have identical error bars, these fits will be equally weighted between the $q$-bins at $|q|\leq q_{\mathrm{max}}/2$ and those with $|q|<q_{\mathrm{max}}/2$, i.e., there will be an equal number $N_q/2$ of $q$-bins to fit in each interval.
		For a simultaneous three-dimensional fit, on the other hand, the same separation of the $q$-bins now yields $(N_q/2)^3 = N_q^3/8$ $q$-bins with $|q|<q_{\mathrm{max}}/2$ in each direction, and $N_q^3 - N_q^3/8 = 7 N_q^3/8$ $q$-bins which are outside this region.  The three-dimensional fit must therefore fit a proportionately much larger number of $q$-bins at large, off-axis values of $q$ than at small, on-axis values of $q$.  Thus, as we will see below, three-dimensional fits will tend to better represent large-$q$ structure of the correlation function, while one-dimensional fits will tend to represent the correlation function more closely near the origin in $q$-space.  Since the smallest lengthscales in the system generate the widest structures in $q$-space, this implies that three-dimensional fits will tend to yield \textit{smaller} estimates for the HBT radii than one-dimensional fits.
		\item \textit{Performing fit-range studies.}  One method commonly used for testing the convergence of a fit to a correlation function involves varying the $q_{\mathrm{max}}$ of the bins which are used in the fit \cite{Frodermann:2006sp}.  Varying $q_{\mathrm{max}}$ in this way constitutes a fit-range study.  Conducting such a study allows one to explore how the quality of the fit is affected by where the $q$-bins are cut off.  If $q_{\mathrm{max}}$ is too small, then the fit will over-represent the shape of the correlation function at the $q$-origin, and under-represent its shape at large-$q$.  An adequate fit to the entire correlation function must therefore be stable with respect to choice of $q_{\mathrm{max}}$.  Because of the difficulty of computing the correlation function at the large number of points required to reliably perform fit-range studies, we currently have not implemented this feature in our analysis.
		\item \textit{Including experimental uncertainties.}  Estimation and incorporation of systematic and statistical uncertainties form an extremely intricate and involved component of experimental HBT analyses.  In general, different $q$-bins are subject to different levels of uncertainty, and this uncertainty directly affects the quality of the fit which one extracts from the correlation function.  In order to provide a meaningful comparison between the theoretically computed correlation functions and those measured experimentally, it is essential to correctly account for the presence of statistical uncertainties, especially when the measured correlation function deviates significantly from a Gaussian form.
	\end{enumerate}
	After computing the correlation function \eqref{CFWR_defn}, we extract the GF HBT radii from it by performing a least-squares fit to the form \eqref{corrfunc_functional_defn}.  To do this we minimize the $\chi^2$-function for the correlation function, which we define by
	\begin{equation}
		\chi^2 \equiv \sum^{N}_{k=1} \l[ \frac{C(\vec{q}^{(k)},\vec{K}) - C_{\mathrm{fit}}(\vec{q}^{(k)},\vec{K})}{\sigma_k} \r]^2,
		\label{chisq_fcn}
	\end{equation}
	where $C(\vec{q}^{(k)},\vec{K})$ is the computed value of the correlation function in the $k$th $\vec{q}$-bin, the index $k$ ranges over the $N$ total points (i.e., bins) for which $C(\vec{q},\vec{K})$ has been computed.
	
	For the results presented in this paper, we take $\sigma_k = 10^{-3}$ for all $k$.\footnote{
	$\sigma_k$ is a placeholder for the uncertainty of the ``data'' (which in our case are obtained
	from a calculation which, ideally, should have zero uncertainty) to which the functional form
	\eqref{corrfunc_functional_defn} is fitted.
	}
This means that deviations of the fit from the data points in the small-$q$ region (where the correlation function $C(\vec{q}^{(k)},\vec{K})$ is the largest) will make larger contributions to the total $\chi^2$ of the fit than points in the large-$q$ region (with the exception that we omit the point $\vec{q}=0$ from the fit, since it is not experimentally accessible, and its omission has a negligible effect on the fit radii).  Our approach here differs from that adopted in most experimental analyses, which fit the quantity $\ln \l( C(\vec{q}^{(k)},\vec{K})-1 \r)$ instead of $C(\vec{q}^{(k)},\vec{K})$.  If we were to compute HBT radii to be compared to experimental data we would have to follow the experimental procedure.

	The minimization itself is implemented numerically in terms of standard GSL routines designed for this purpose.

	As we have already observed, the fitting of the correlation function is highly sensitive to the distribution of points in $\vec{q}$-space.  In this paper, we choose the grid of points to have a uniform spacing along the $q_x$, $q_y$, and $q_z$ axes (the corresponding $osl$ coordinates of any $\vec{q}$-point are then obtained with a positive rotation around the $z$ axis by angle $\Phi_K$ according to Eqs. \eqref{osl_xyz_transf}).  After computing the correlation function for each of the $N$ $\vec{q}$-bins, we perform a full, three-dimensional fit of \eqref{CFWR_defn} to \eqref{corrfunc_functional_defn} by minimizing the $\chi^2$-function \eqref{chisq_fcn}. Here we do not attempt to perform fit-range studies or mimic experimental error bars in our fitting procedure.  This is mostly because we find that resonance decays introduce strong non-Gaussianity in the correlation function such that the HBT radii extracted from a Gaussian fit depend sensitively on the binning of the correlation function in $\vec{q}$-space which must therefore be closely matched between theory and experiment for meaningful comparisons.
	
	%%%%%%%%%%%%%%%%%%%%%%%%%%%%%%%%%%%%%%%%%%%%%%%%%%%%%%%
	\section{Results}
	\label{Sec3}
	%%%%%%%%%%%%%%%%%%%%%%%%%%%%%%%%%%%%%%%%%%%%%%%%%%%%%%%
	
%
%%%%%%%%%%%%%%%% Fig. 2  %%%%%%%%%%%%%%%%%%%%
	\begin{figure*}[!htbp]
		\centering
		\includegraphics[width=0.75\textwidth]{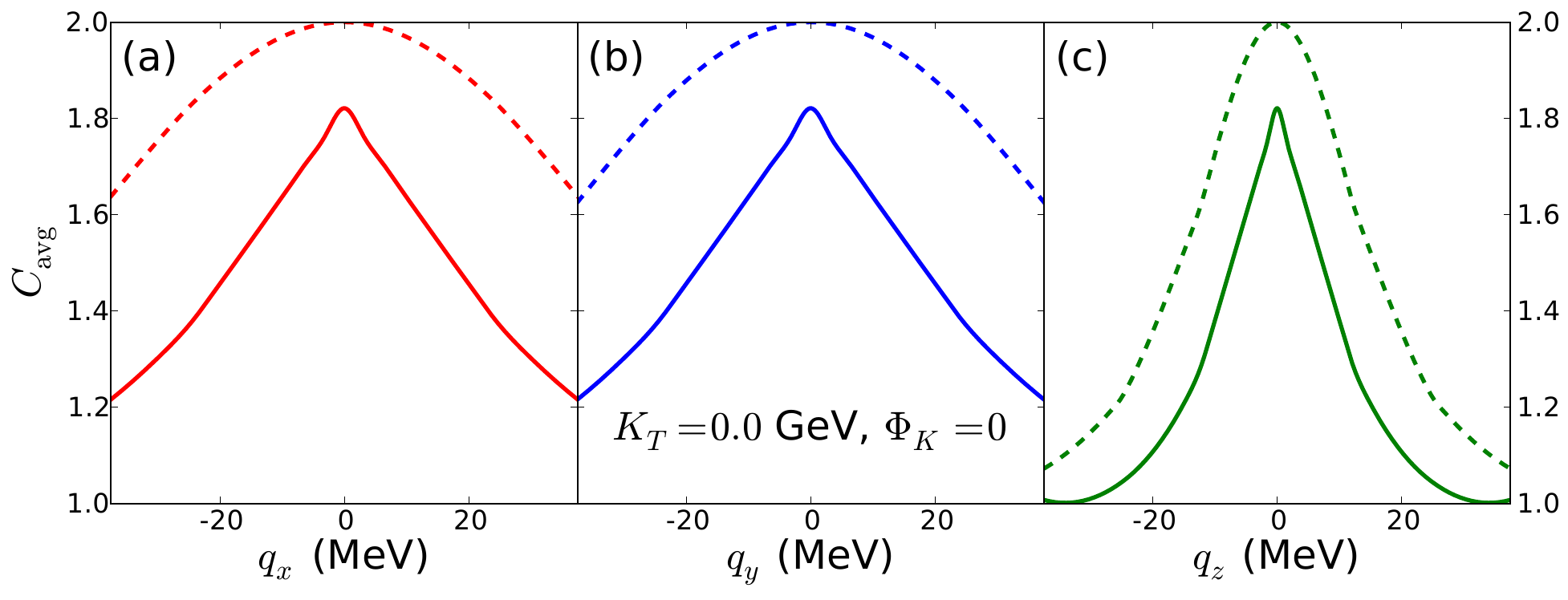}\\
		\includegraphics[width=0.75\textwidth]{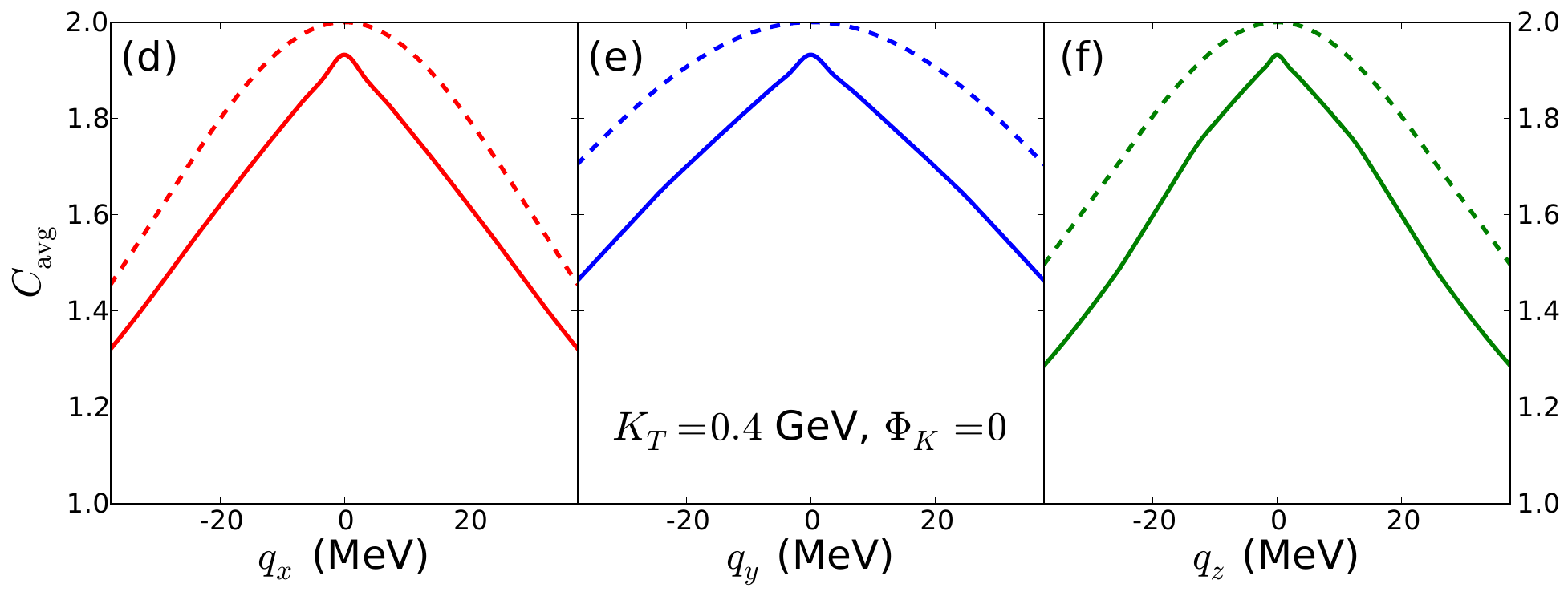}\\
		\includegraphics[width=0.75\textwidth]{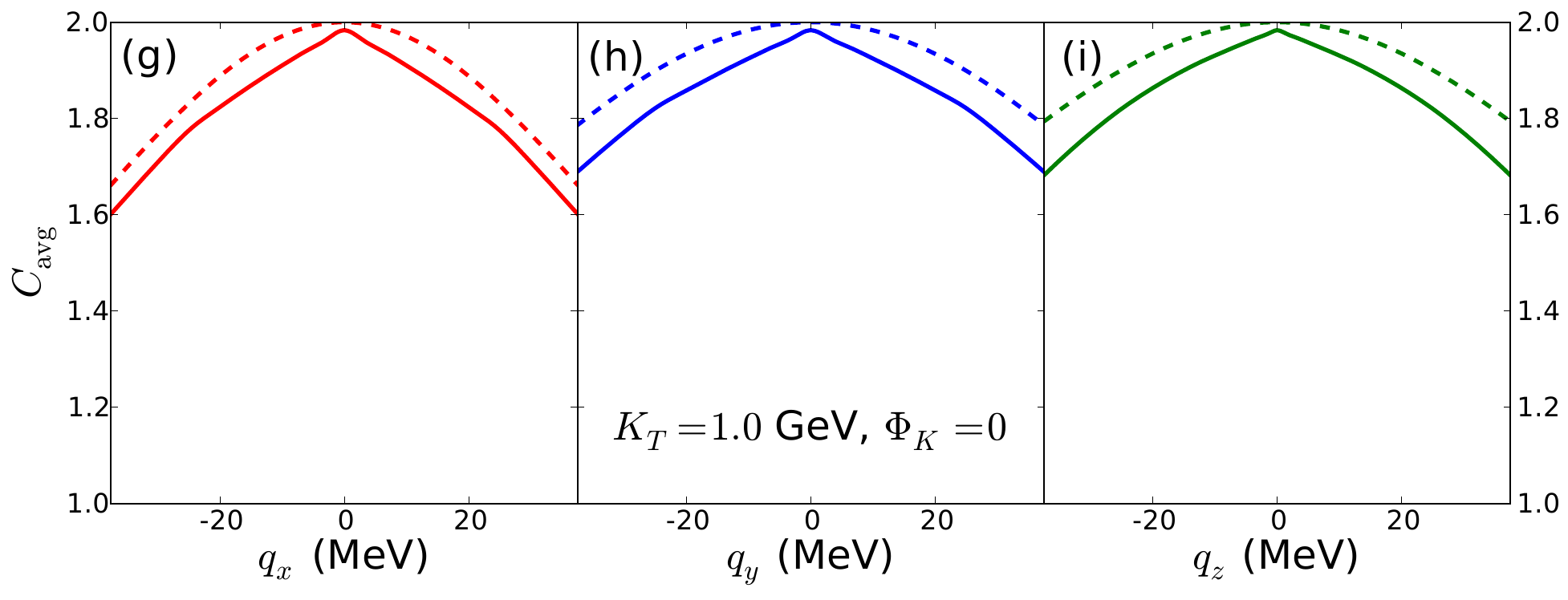}
		\caption{(Color online)
		Slices of the full ensemble-averaged correlation function $C_{\mathrm{avg}}$ 
		\eqref{corrfuncEAdefn} including all resonance decays (solid lines) along the $q_o$, 
		$q_s$ and $q_l$ axes (from left to right), compared with the analogous results for 
		directly emitted (``thermal") pions only (dashed lines), for three choices of the pair
		momentum, $K_T=0,\ 0.4,$ and 1.0\,GeV (from top to bottom). The pair momentum
		$\vec{K}_T$ was chosen to point in x-direction ($\Phi_K=0$) such that $q_x=q_o$ and
		$q_y=q_s$. The ensemble consists of $N_{\ev} = 1000$ hydrodynamically evolved 
		central (0-10\% centrality) Au-Au collisions with $\eta/s=0.08$ at $\sqrt{s}=200\,A$\,GeV.
		\label{CF_w_v_wo_res_comparison}}
	\end{figure*}
%%%%%%%%%%%%%%%%%%%%%%%%%%%%%%%%%%%%
%

	Using the iEBE-VISHNU package \cite{Shen:2014vra} we generated an ensemble of $N_{\ev}=1000$ central (0-10\%) Au+Au events at $\sqrt{s}= 200\,A$\,GeV, and then used the HoTCoffeeh code presented in this paper to compute the HBT correlation functions and radii for pion pairs, using both the SV and GF methods for comparison. The hydrodynamic event sample is identical with the one described and studied in   Ref.~\cite{Plumberg:2015aaw}; it assumes viscous fluid dynamic evolution of the hot matter created in the collision with a constant specific shear viscosity $\eta/s = 0.08$. In this section we study in detail all systematic features of the HBT radii associated with this hydrodynamic event ensemble and compare our results qualitatively with those from earlier studies of more schematic model sources \cite{Wiedemann:1996ig} and of ideal fluid dynamical simulations of smooth initial conditions \cite{Frodermann:2006sp}.

	%%%%%%%%%%%%%%%%%%%%%%%%%%%%%%%%%%%%%%%%%%%%%%%%%%%%%%%
	\subsection{Correlation functions with and without resonance decays}
	\label{Sec3a}
	%%%%%%%%%%%%%%%%%%%%%%%%%%%%%%%%%%%%%%%%%%%%%%%%%%%%%%%
	
	To build intuition for the qualitative influence of resonance decay contributions on the shape of the two-pion correlation function and the HBT radii associated with it, we compare in Fig.~\ref{CF_w_v_wo_res_comparison} the correlation functions for directly emitted pions (dashed lines) with those obtained from the full emission function including all resonance decay contributions (solid lines).

While the correlation functions for directly emitted pions look pretty Gaussian (although a more quantitative analysis exposes that this not really true along the $q_l$ direction \cite{Wiedemann:1996ig,Frodermann:2006sp}), adding the contributions from resonance decays clearly distorts the shape of correlation function in all three directions, making it much sharper than a Gaussian near $q=0$. In addition, the peak of the correlation function at $q=0$ never reaches the value 2 once resonance decay pions are included, on account of the long-lived resonances such as the $\eta$ meson which contribute to the pion yield in the denominator of the correlation function but whose contribution to the numerator is almost a $\delta$-function at $q=0$ and cannot be resolved experimentally, due to finite momentum resolution.\footnote{
	Another reason for suppressing the peak of experimentally measured correlation functions 
	below the value of 2, not studied in the present paper, could be a violation of the assumption 
	of independent particle emission, e.g. through phase coherence among the emitted pions 
	\cite{Andreev:1992pu, Heinz:1997mr, Csorgo:1998tn, Gangadharan:2015ina}.
	}
%
%
%%%%%%%%%%%%%%%%%%%% Fig. 3 %%%%%%%%%%%%%
	\begin{figure}[!htbp]
		\centering
		\includegraphics[width=0.8\linewidth]{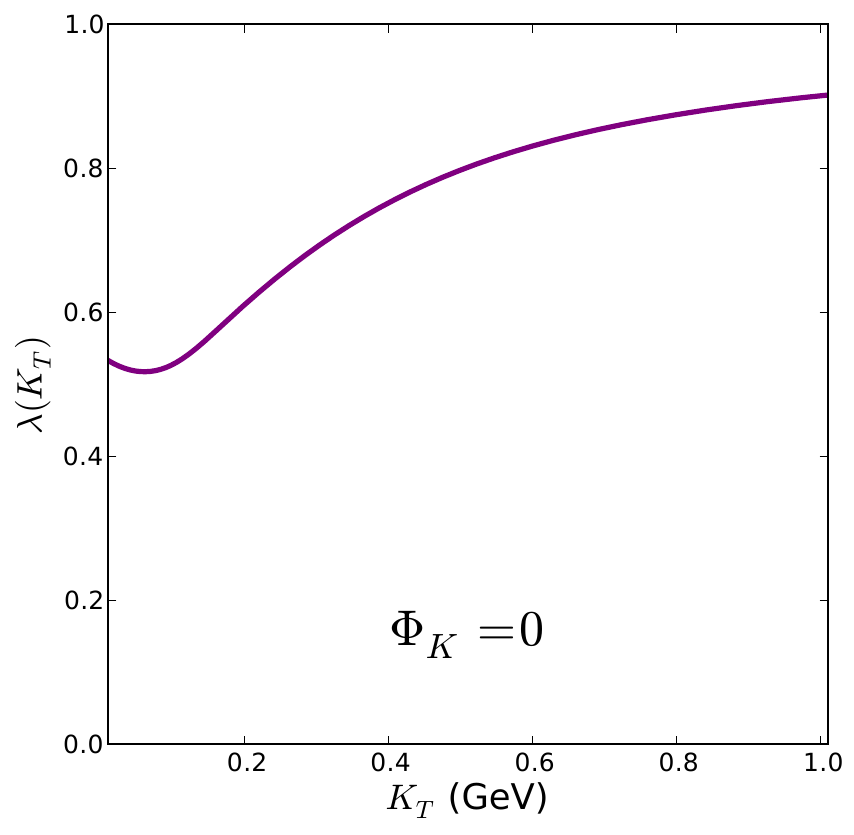}
		\caption{(Color online)
		The intercept parameter $\lambda(\vec{K})$ (defined in Eq.~\eqref{corrfunc_functional_defn}) 
		as a function of $\vec{K}$, for a three-dimensional Gaussian fit to the full correlation function  
		including resonances shown in Fig.~\ref{CF_w_v_wo_res_comparison}.
		\label{lambdas_vs_KT}}
	\end{figure}
%%%%%%%%%%%%%%%%%%%%%%%%%%%%%%%%%%%%
%	
Both effects, the depression of the correlation peak at $q=0$ and the non-Gaussian distortion of the $\vec{q}$-dependence, are stronger for pion pairs with small pair momentum $K$ and slowly die out at large pair momentum. For the intercept $\lambda$, extracted as an additional fit parameter in \eqref{corrfunc_functional_defn}, this is shown in Fig.~\ref{lambdas_vs_KT} (see also \cite{Wiedemann:1996ig}). This reflects the fact \cite{Hagedorn:1965st} that the decay phase-space favors low transverse momenta for decay pions from heavy resonances while at large transverse momenta the directly emitted pions dominate. Radial flow reduces this bias \cite{Wiedemann:1996ig} but does not fully eliminate it. Furthermore, as noted in \cite{Frodermann:2006sp}, $\lambda(\vec{K})$ may continue to deviate from unity even if resonances decays are excluded, due to the inability of a three-dimensional Gaussian fit to fully capture the non-Gaussian $q_l$ dependence that survives even for thermally emitted pions due to the boost-invariant longitudinal expansion of the source \cite{Wiedemann:1996ig}.

	%%%%%%%%%%%%%%%%%%%%%%%%%%%%%%%%%%%%%%%%%%%%%%%%%%%%%%%
	\subsection{HBT radii including resonance decays: SV method}
	\label{Sec3b}
	%%%%%%%%%%%%%%%%%%%%%%%%%%%%%%%%%%%%%%%%%%%%%%%%%%%%%%%
	
	In this subsection we study the HBT radii extracted via the SV methods, their event-by-event distributions, means and variances, for the same ensemble of 1000 events discussed above.
	
\begin{figure*}[!htbp]
		\centering
		\includegraphics[width=0.75\textwidth]{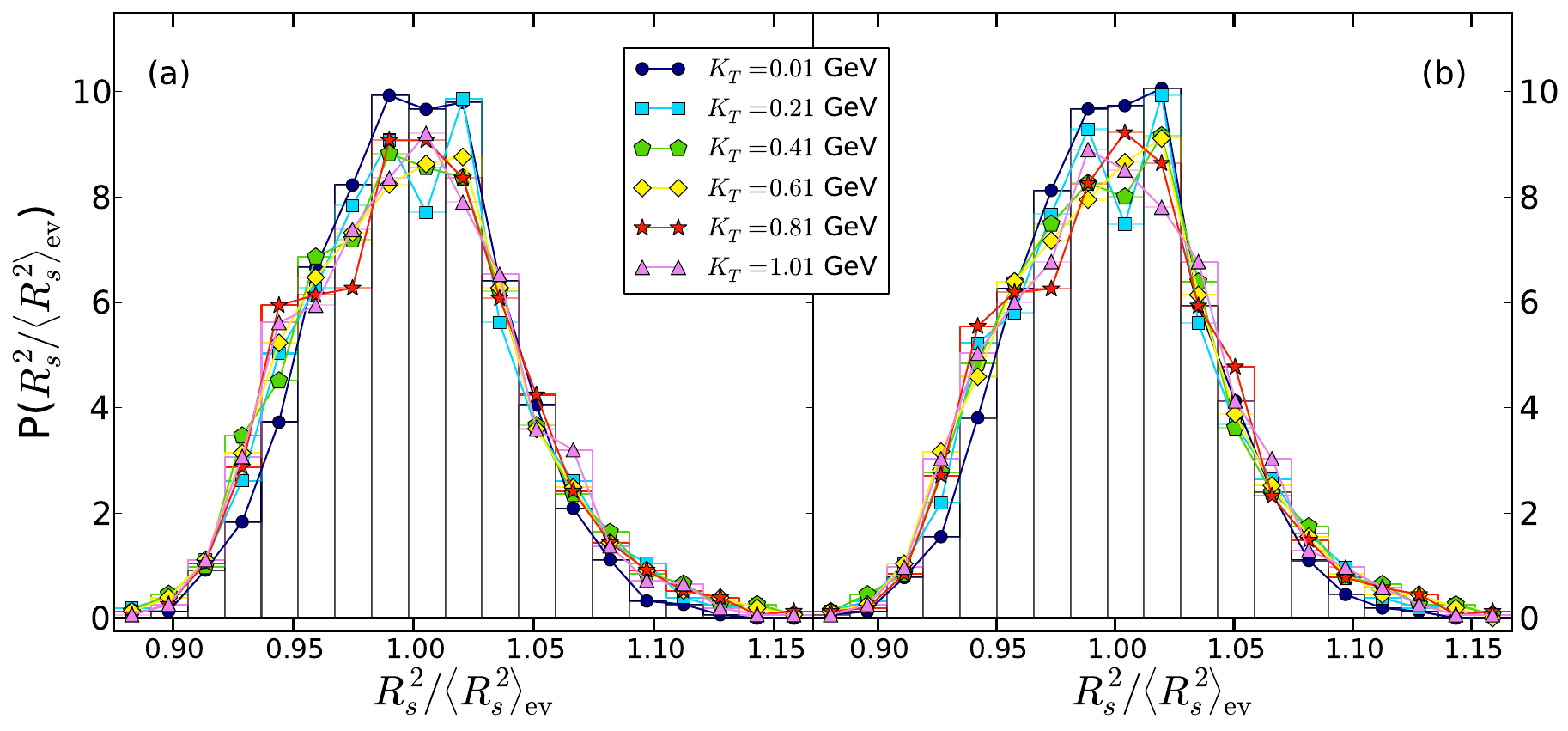}
		\caption{(Color online)
		Event-by-event distributions of the azimuthally averaged SV $R^2_{s,0}$ 
		\cite{Plumberg:2013nga} (denoted simply as $R_s^2$ in the figure), (a) with 
		and (b) without the $\delta f$ correction. 
		\label{EBE_SV_R2s_dist}}
	\end{figure*}

\begin{figure*}[!htbp]
		\centering
		\includegraphics[width=0.75\textwidth]{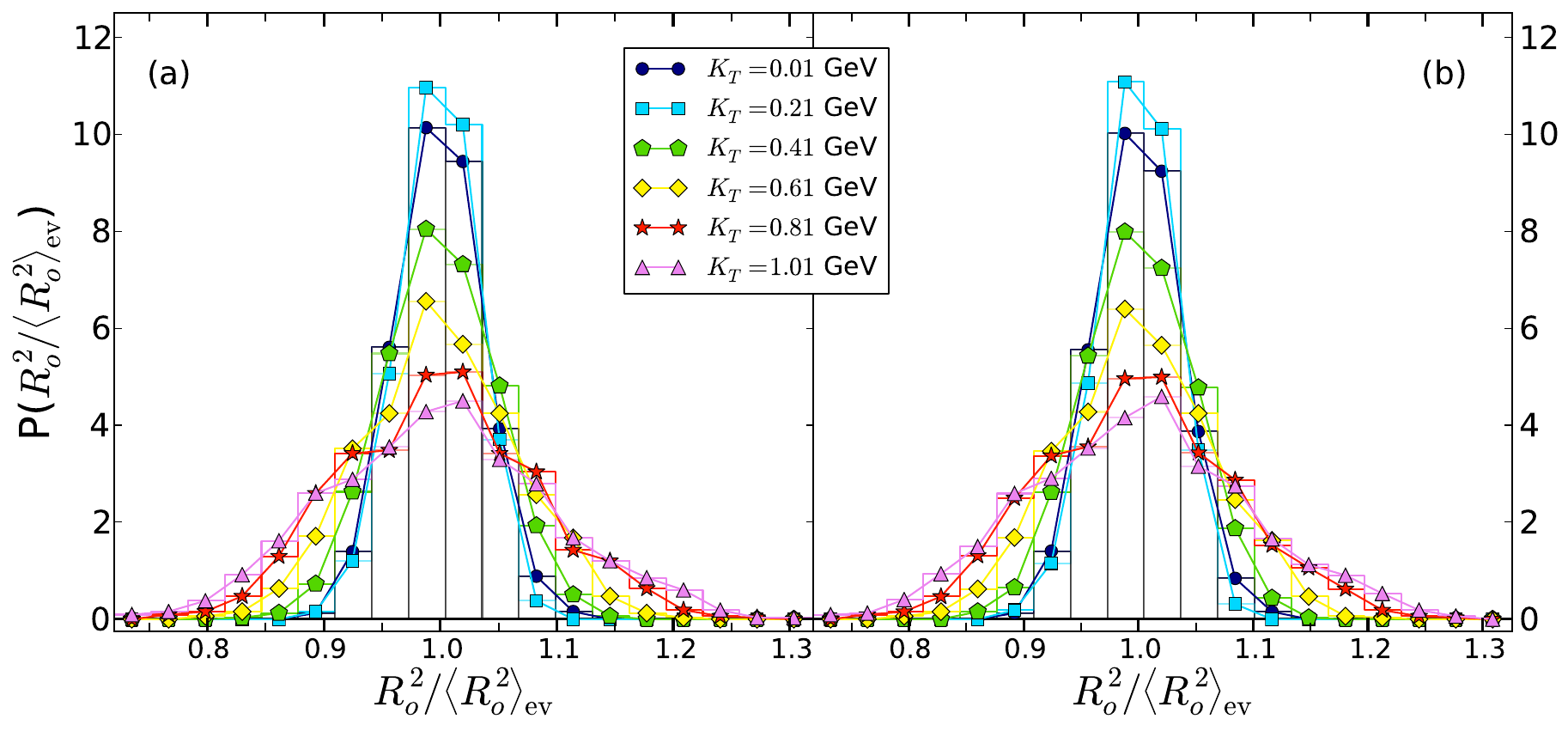}
		\caption{(Color online)
		Same as Fig.~\ref{EBE_SV_R2s_dist}, but for the outward radius parameter SV $R^2_{o,0}$.
		\label{EBE_SV_R2o_dist}
		}
	\end{figure*}

\begin{figure*}[!htbp]
		\centering
		\includegraphics[width=0.75\textwidth]{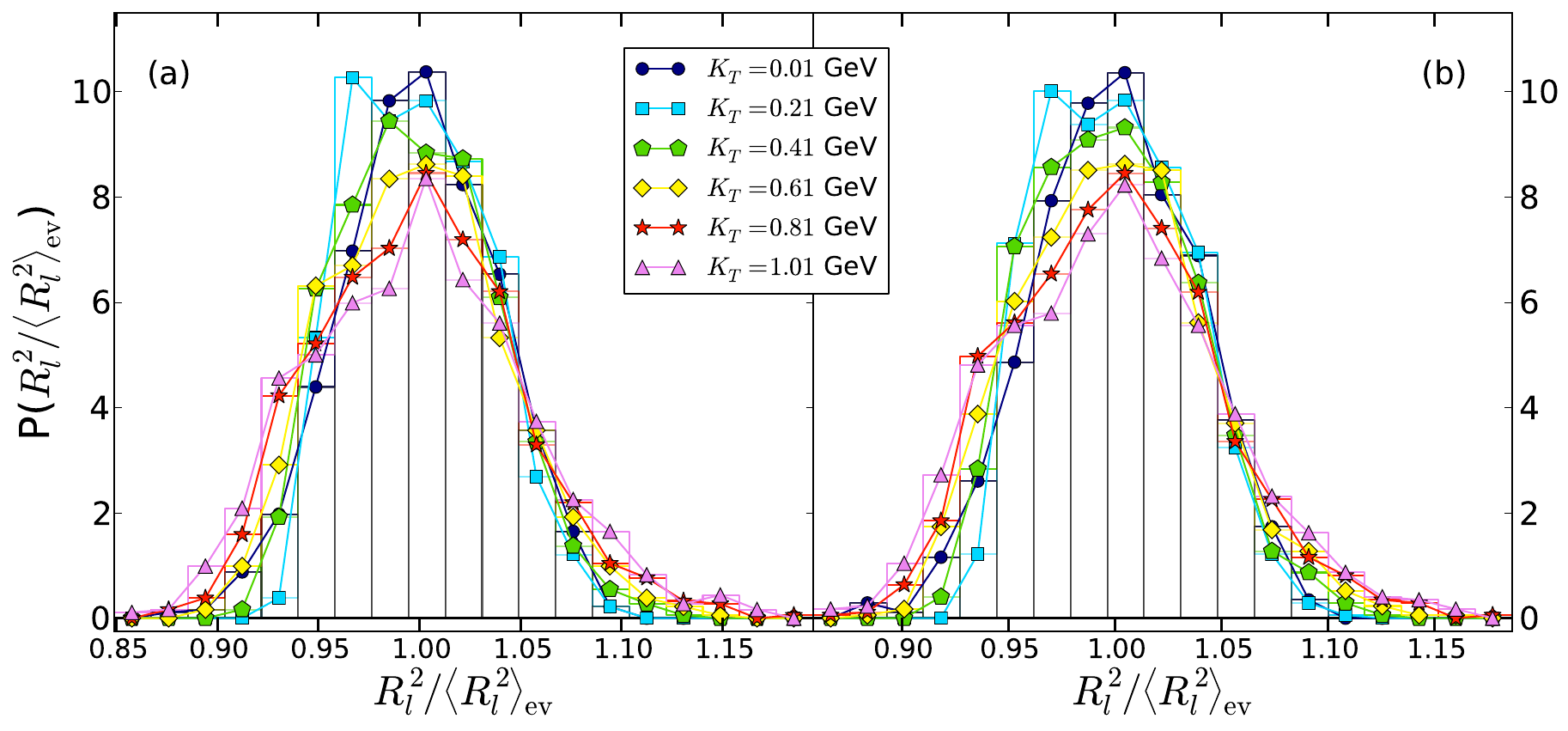}
		\caption{(Color online)
		Same as Fig.~\ref{EBE_SV_R2s_dist}, but for the longitudinal radius parameter SV $R^2_{l,0}$.
		\label{EBE_SV_R2l_dist}}
	\end{figure*}

	Figures~\ref{EBE_SV_R2s_dist} -- \ref{EBE_SV_R2l_dist} show the event-by-event distributions of the $\Phi_K$-averaged sideward, outward and longitudinal radius parameters, respectively, normalized by their mean values, for six different values of the pair momentum $K_T$. The left panels (a) show the full result, the right panels (b) are obtained by removing the viscous $\delta f$ correction from Eq.~(\ref{cooper_frye_defn2}). We see that $\delta f$ has no obvious visible effect on these distributions. In the rest of this paper we therefore always include the $\delta f$ correction.

	%%%%%%%%%%%   Fig. 4 %%%%%%%%%%%%%%%%%%%%%%%%%
	%%%%%%%%%%%%%%%%%%%%%%%%%%%%%%%%%%%%%%%%
	%
	%%%%%%%%%%%%%%% Fig. 5 %%%%%%%%%%%%%%%%%
	%%%%%%%%%%%%%%%%%%%%%%%%%%%%%%%%%%%%
	%
	%%%%%%%%%%%  Fig. 6 %%%%%%%%%%%%%%%%%%%%%%%%%
	%%%%%%%%%%%%%%%%%%%%%%%%%%%%%%%%%%%%
	%

	The shapes of these distributions show very little $K_T$-dependence for the sideward and longitudinal radii, with a $\sim10\%$ width over the entire range of $K_T$ values studied. As seen in Fig.~\ref{Fig:SV_R2ij0_DEA_0_08} below, the mean values of both $R_s^2$ and $R_l^2$ decrease with increasing pair momentum; Figs.~\ref{EBE_SV_R2s_dist} and \ref{EBE_SV_R2l_dist} imply that the widths of their distributions decrease in sync with their mean values.
	
	The similarly normalized width of the outward radius parameter $R^2_o$, shown in Fig.~\ref{EBE_SV_R2o_dist}, strongly increases with increasing pair momentum, doubling from about 15\% at small $K_T$ to more than 30\% at $K_T=1$\,GeV. As discussed in Refs.~\cite{Plumberg:2015eia,Plumberg:2015aaw}, this increased variance at higher $K_T$ can be attributed to the contribution to $R_o^2$ from the emission duration, $\beta_T^2\bigl(\langle t^2\rangle{-}\langle t\rangle^2\bigr)$, which strongly fluctuates at large $K_T$. These increasing fluctuations of the emission duration are generic and occur whether or not resonance decay contributions are included. We will see below that the GF HBT radii exhibit the same feature.
			
	We next consider the result of ensemble averaging the SV HBT radii, including all resonance decay contributions. We present these results in Fig.~\ref{Fig:SV_R2ij0_DEA_0_08}.
	%
	%%%%%%%%%%%%%% Fig. 7 %%%%%%%%%%%%%%%%%%%%%%
	\begin{figure}[!htbp]
		\centering
		\includegraphics[width=0.95\linewidth]{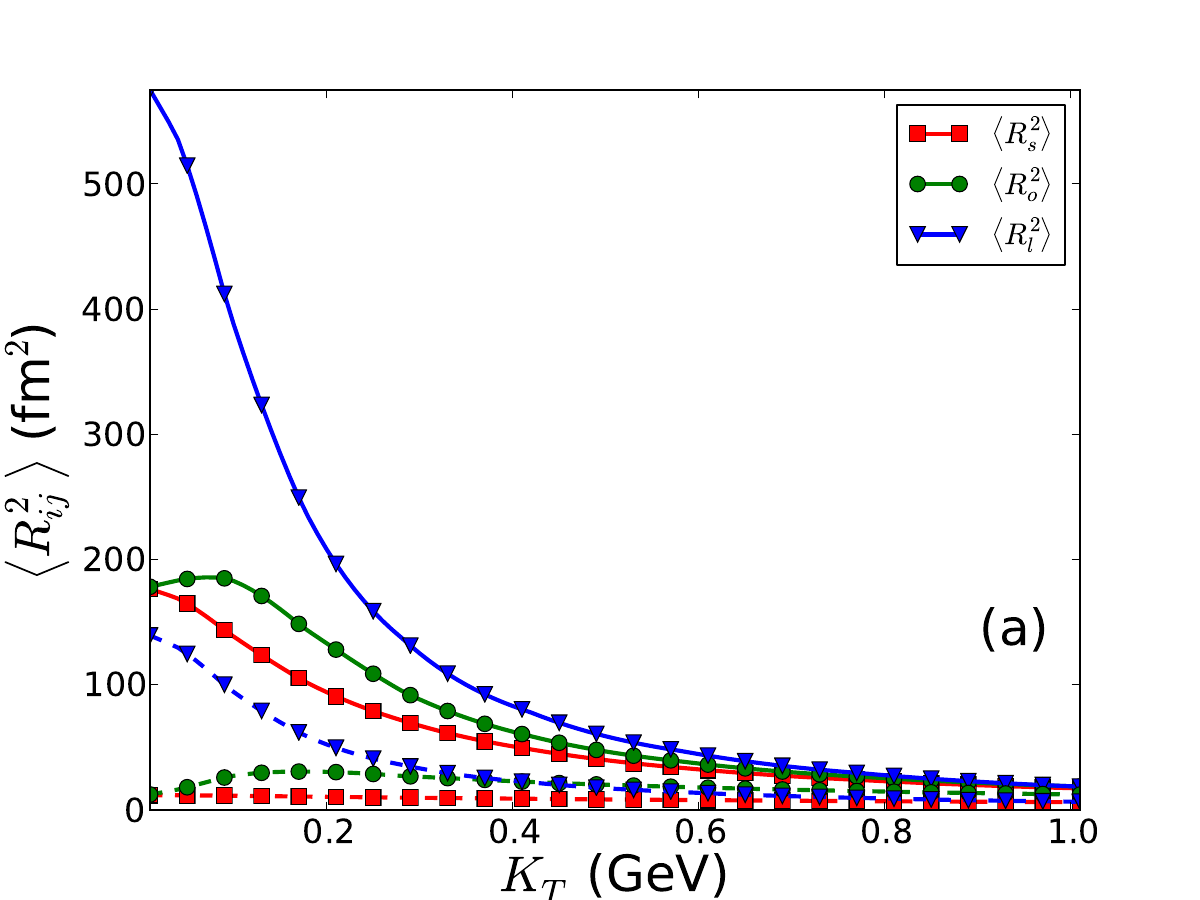}\\
		\includegraphics[width=0.95\linewidth]{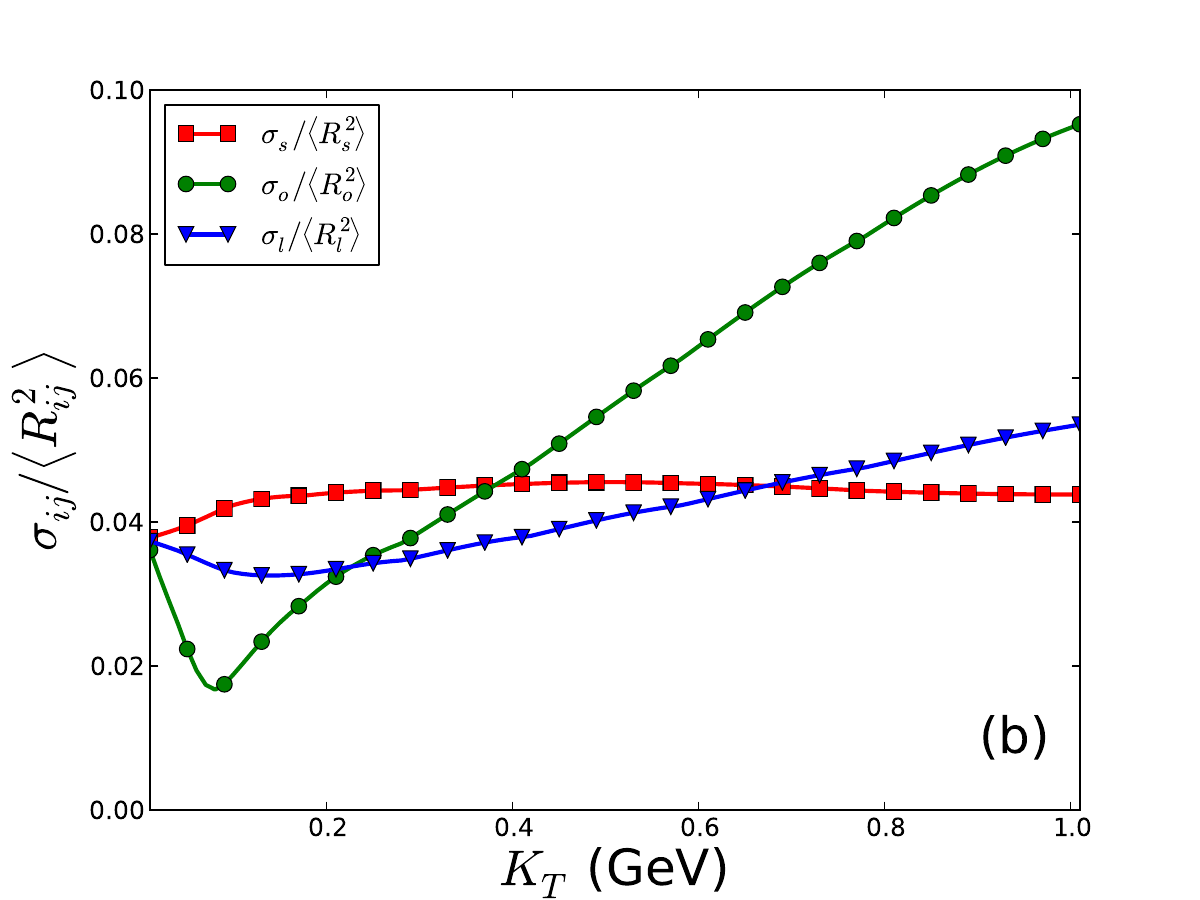}
		\caption{(Color online)
		The azimuthally and ensemble averaged SV HBT radii (a) and their normalized variances 
		(b) as a function of pair momentum $K_T$, including all resonance decays, for the central
		Au-Au collision events studied in this paper. In panel (a) solid lines show results that include all 
		resonance decay contributions while dashed lines show the HBT radii for only the directly
		emitted pions.
		\label{Fig:SV_R2ij0_DEA_0_08}
		}
	\end{figure}
	%%%%%%%%%%%%%%%%%%%%%%%%%%%%%%%%%%%%
	%
	As discussed in Ref.~\cite{Plumberg:2015mxa}, the HBT radii corresponding to the ensemble-averaged correlation function \eqref{corrfuncEAdefn} do not agree with the direct arithmetic average of HBT radii from the individual fluctuating events, but differ by an event multiplicity weight. We checked that for the ensemble of events studied here multiplicity fluctuations are small and the difference between the two definitions of the average HBT radius parameters is less than 1\%. We therefore present only the arithmetically averaged HBT radius parameters $\langle R_i^2\rangle = \sum_{k=1}^{N_\mathrm{ev}} (R_i^2)^{(k)}/N_\mathrm{ev}$.

	Figure~\ref{Fig:SV_R2ij0_DEA_0_08}a shows as solid lines the azimuthally and ensemble averaged sideward, outward and longitudinal radii from the SV method, including all resonance decays, as a function of pair momentum $K_T$. They are very much larger than those obtained from the emission function for just the directly emitted pions (dashed lines, see also \cite{Plumberg:2015mxa}). For the squared transverse radii at $K_T=0$ the difference is a factor 15, corresponding to radii that are almost a factor 4 larger. This is an artifact of the SV method which measures the curvature of the two-pion correlation function at $q=0$ rather than its inverse width. For small $K_T$ this curvature is large, as seen in the top row of Fig.~\ref{CF_w_v_wo_res_comparison}, being dominated by the very large emission regions and emission durations of pions from the longest-lived resonances in the mix. For larger $K_T$ values pions from resonance decays play a less important role, and the difference between the curvature of the correlation function at $q=0$ and its inverse width becomes less pronounced. Generally speaking we see, however, that SV HBT radii (which measure the curvature of the correlation function at $q=0$) are a poor way of characterizing its shape (in particular, its inverse width) once resonance decays are taken into account, especially for pion pairs with small to moderate pair momentum.
	
Figure~\ref{Fig:SV_R2ij0_DEA_0_08}b shows the normalized variances (relative widths) of the event-by-event distribution of the SV HBT radii. By comparing with Refs.~\cite{Plumberg:2015mxa,Plumberg:2015aaw} (c.f. Fig.~1 in \cite{Plumberg:2015aaw}, for instance) we observe that resonance decay contributions lead to a slight reduction of these normalized variances. This is easily understood: the variances of the HBT radii associated with the emission regions of decay pions (which reflect the fluctuations in the emission regions of their thermally emitted parent resonances) are expected to be similar to those of the thermally emitted pions and not to increase at the same rate as their mean values as the resonance lifetimes increase. Indeed, we observe that the fluctuations of the source variances including resonance decay contributions shown in Fig.~\ref{Fig:SV_R2ij0_DEA_0_08}b show qualitative similarity with the same fluctuations when resonances are excluded \cite{Plumberg:2015mxa,Plumberg:2015aaw}. Most notably, while the widths of $R^2_s$ and $R^2_l$ stay relatively constant when increasing $K_T$, the normalized variance of $R_o$ strongly grows with increasing $K_T$; this reflects the broadening of the $R_o^2$ distribution seen in Fig.~\ref{EBE_SV_R2o_dist}.
	
	%%%%%%%%%%%%%%%%%%%%%%%%%%%%%%%%%%%%%%%%%%%%%%%%%%%%%%%
	\subsection{HBT radii including resonance decays: GF method}
	\label{Sec3c}
	%%%%%%%%%%%%%%%%%%%%%%%%%%%%%%%%%%%%%%%%%%%%%%%%%%%%%%%
	
	%%%%%%%%%%%%% Fig. 8 %%%%%%%%%%%%%%%%%%%%%%%
	\begin{figure*}[!htbp]
		\centering
		\includegraphics[width=0.32\textwidth]{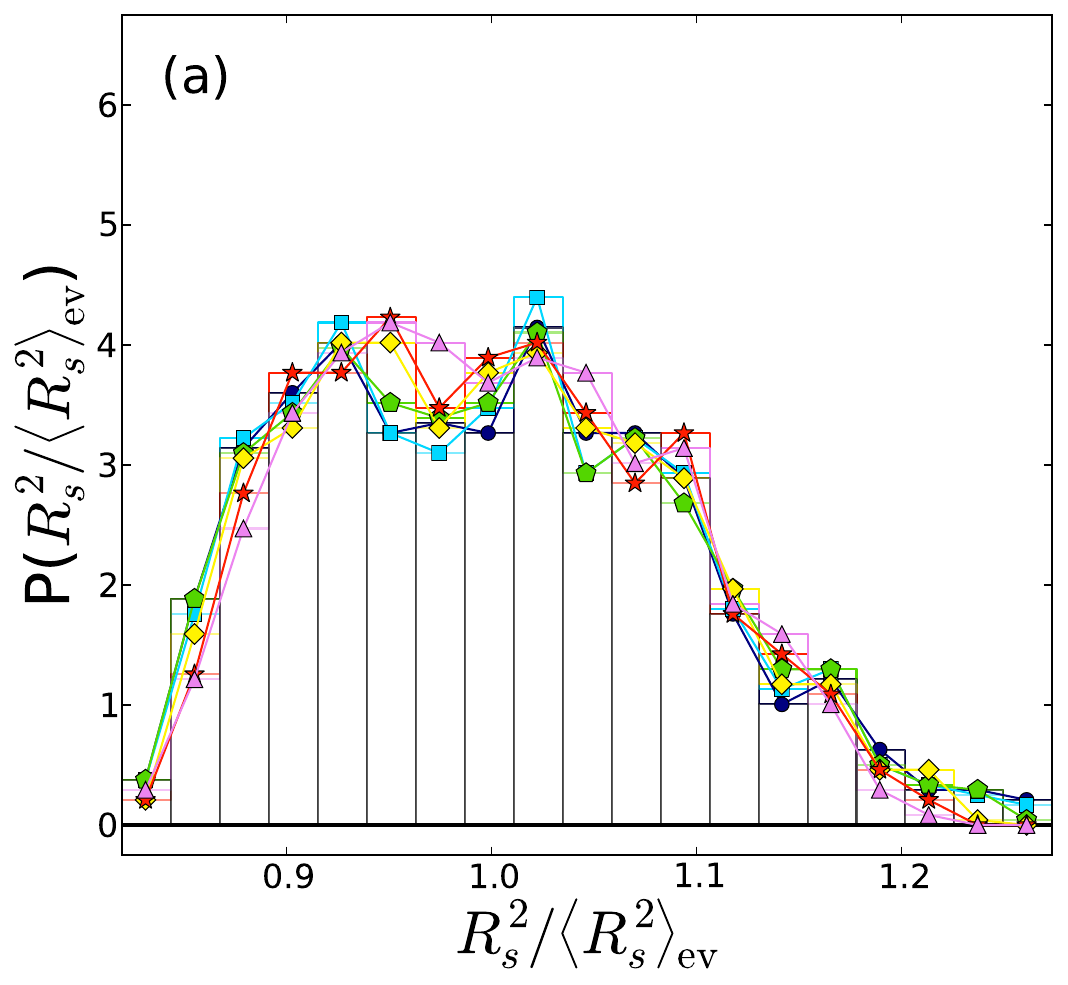} 
		\includegraphics[width=0.32\textwidth]{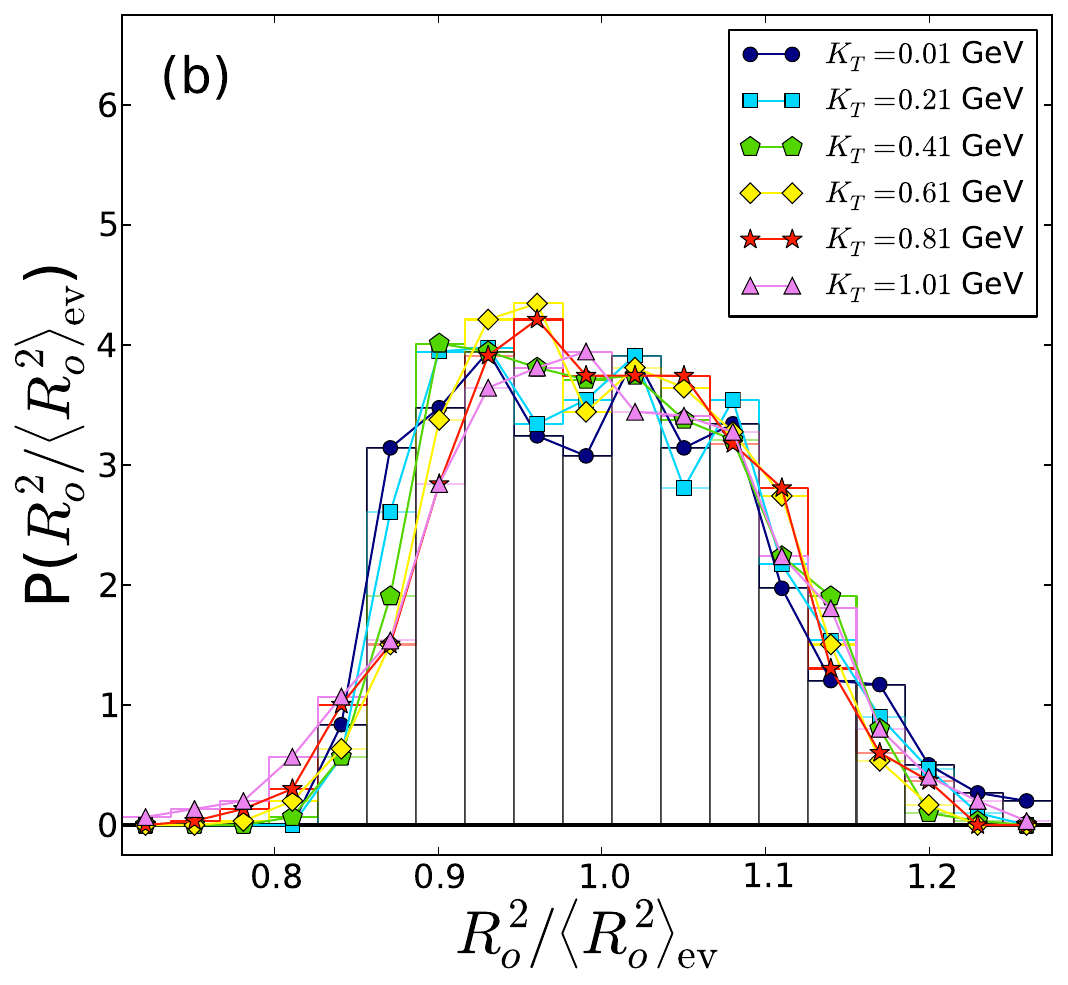} 
		\includegraphics[width=0.32\textwidth]{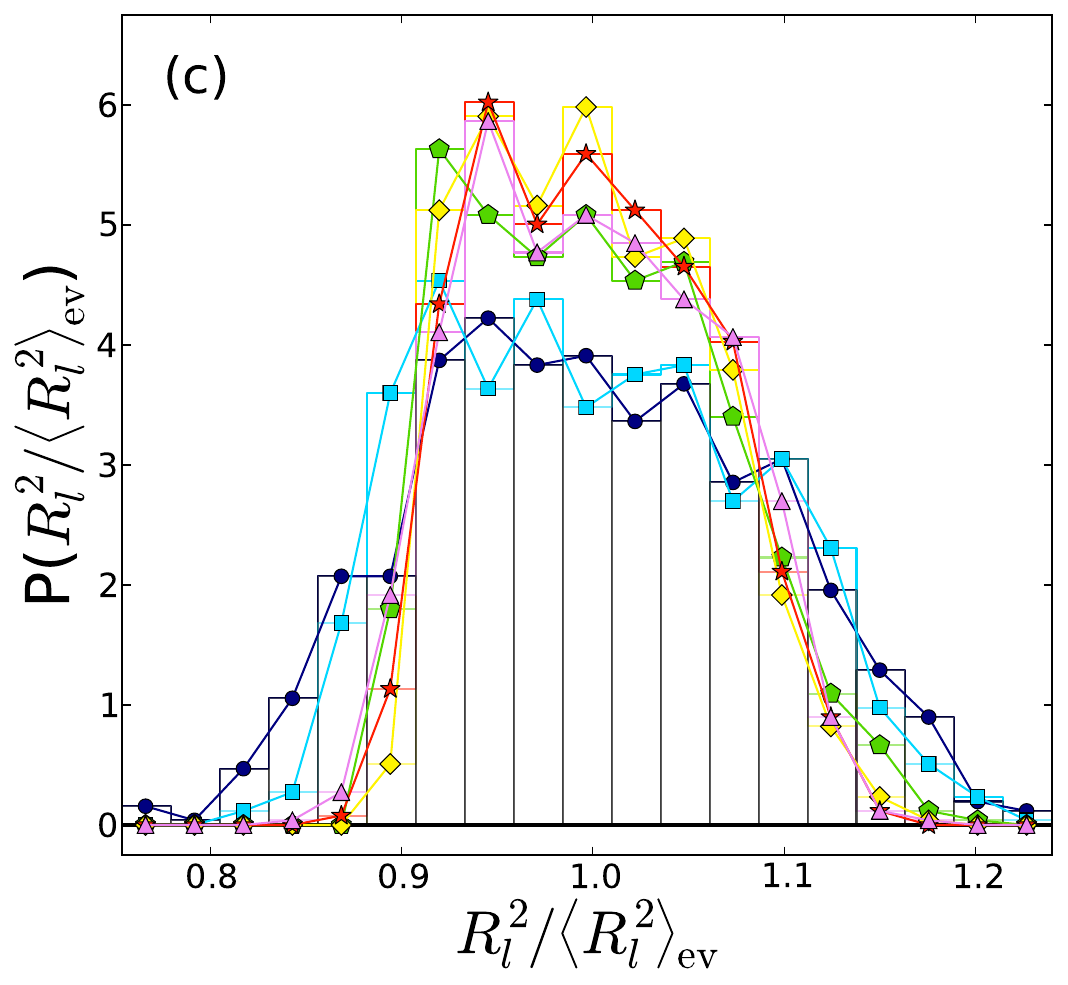}
		\caption{Event-by-event distributions of GF HBT radii, for several different values of $K_T$.
		\label{EBE_HBT_GF}}
	\end{figure*}
	%%%%%%%%%%%%%%%%%%%%%%%%%%%%%%%%%%%%%%%%
	%
	We now contrast these results for the corresponding ones with the GF method of computing the HBT radii. The results shown in this subsection were obtained by minimizing the $\chi^2$ of a three-dimensional Gaussian fit, calculated over a grid of $7^3$ points $(q_o,q_s,q_l)$, with $q_s,q_o\in\{0,\pm25.0,\pm50.0,\pm75.0\}$\,MeV and $q_l\in\{0,\pm12.5,\pm25.0,\pm37.5\}$\,MeV, which was subsequently interpolated (``fleshed out" - see the Appendix) to a denser grid of $N=51^3$ points, spaced uniformly over the same region in $\vec{q}$-space. In the following subsection we discuss the sensitivity of the HBT radii extracted from the Gaussian fit to the details of the fit procedure, including fit range and grid point spacing.  
	
	The event-by-event distributions of the GF HBT radii are presented in Fig.~\ref{EBE_HBT_GF}.  We note that the shape of all three distributions shows less variability with $K_T$ than seen in  Figs.~\ref{EBE_SV_R2s_dist}{--}\ref{EBE_SV_R2l_dist} for the SV HBT radii. The relative widths of all three probability distributions is larger than in the SV case and, in particular, the $R^2_o$ distribution shows much less of a width difference between small and large pair momenta.
	
\begin{figure}[!htbp]
		\centering
		\includegraphics[width=0.95\linewidth]{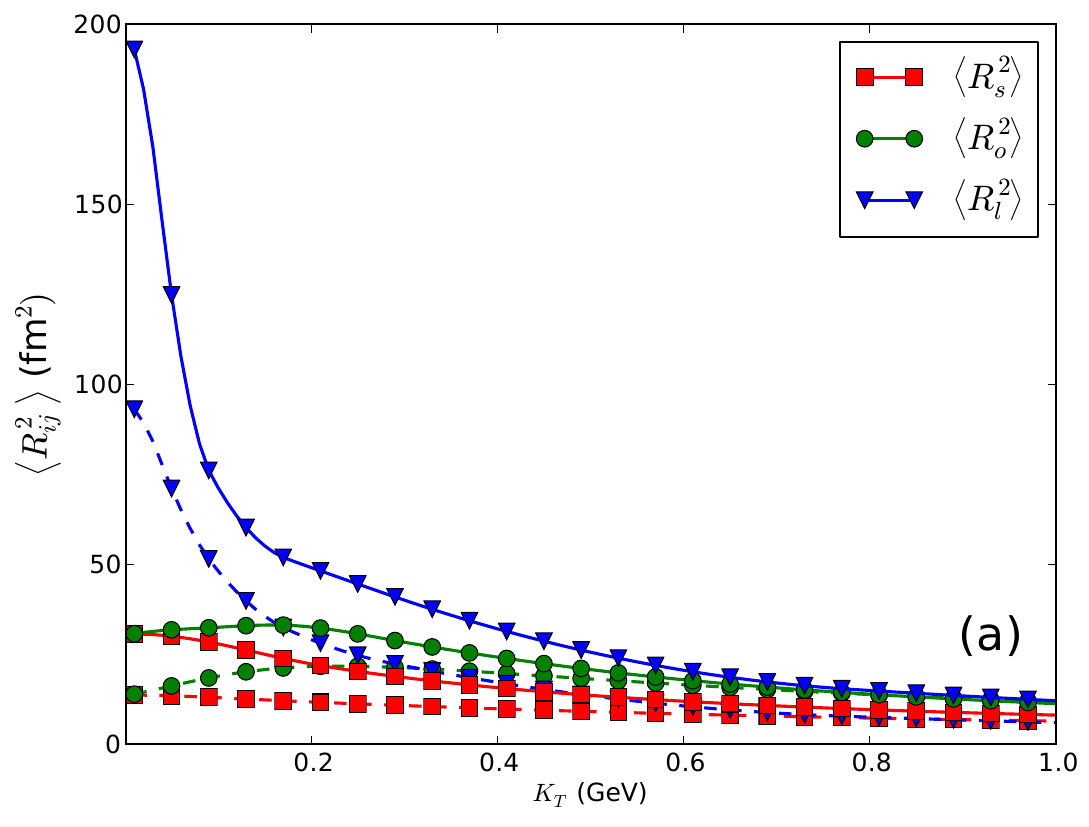}\\
		\includegraphics[width=0.95\linewidth]{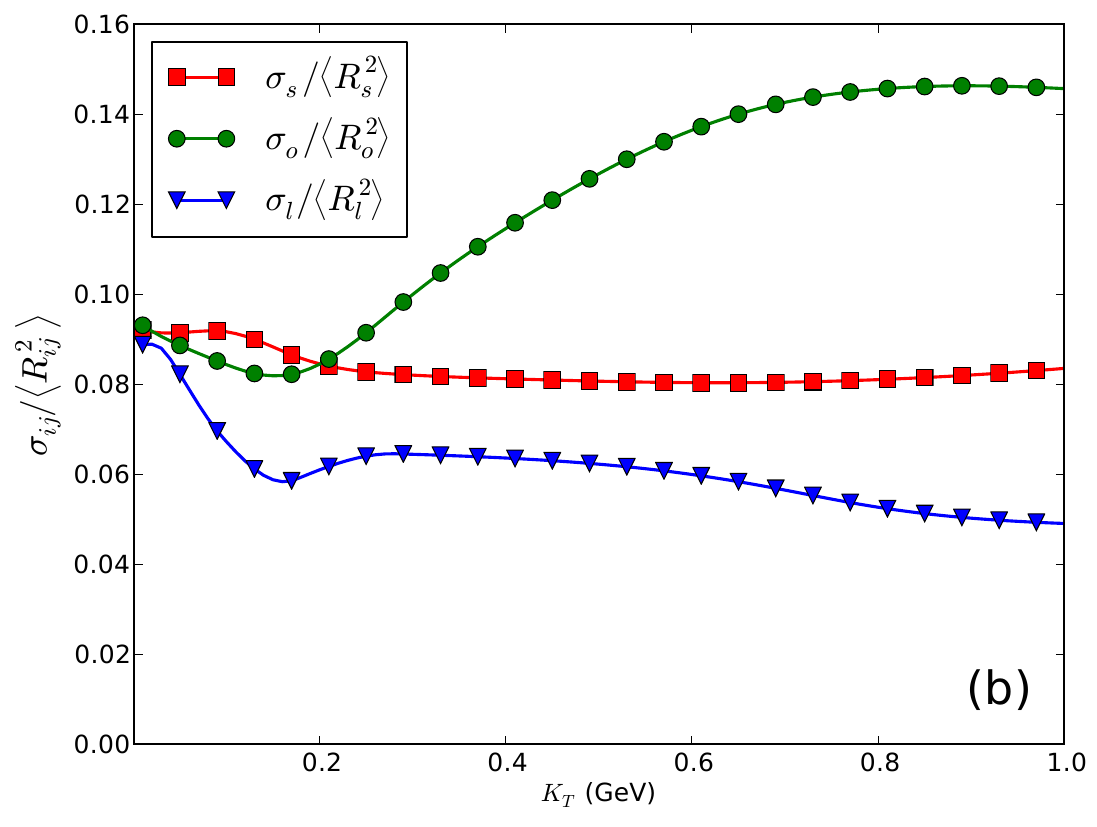}
		\caption{(Color online)
		Same as Fig.~\ref{Fig:SV_R2ij0_DEA_0_08}, but for the mean radii and normalized
		variances of the GF radii.  Solid lines correspond to the radii with all resonances included, while the dashed lines represent the GF radii using thermal $\pi^+$s only.
		\label{EAGFHBT}
		}
	\end{figure}

	The main reason for this can be seen in Fig.~\ref{EAGFHBT}a which shows that the GF HBT radii (by which the relative widths are normalized) are much smaller than the SV radii. A closer view, taking into account the information on the normalized variances shown in Fig.~\ref{EAGFHBT}b, reveals that also the variances of the GF HBT radii are smaller than those of the SV radii, but the larger reduction is seen by the radii themselves: a factor 6 for the squared transverse radii and a factor 3 for the squared longitudinal radius at $K_T=0$.
	%
	%%%%%%%%%%%%% Fig. 9 %%%%%%%%%%%%%%%%%%%%%%%
	%%%%%%%%%%%%%%%%%%%%%%%%%%%%%%%%%%%%
	%
The Gaussian widths of the correlation function are seen to be much less sensitive to the relatively small contribution of very long-lived resonances than the curvature at $q=0$ and are instead dominated by the bulk of pions being emitted either directly or from short-lived resonances. Still, these short-lived resonances significantly increase the GF squared radii of the full emission function (solid lines) over those of the directly emitted pions (dashed lines), by factors 2.5 and 2 for the transverse and longitudinal squared radii, respectively, at $K_T=0$. For the hydrodynamic sources studied here this resonance decay effect on the HBT radii is larger than what was observed in \cite{Wiedemann:1996ig} for a hydrodynamically motivated Gaussian model emission function. 
	
	The pair momentum dependence of the relative widths of the HBT radii distributions are shown in Fig.~\ref{EAGFHBT}b. As for the SV method we see outward radii fluctuations that strongly increase with $K_T$, for the same reason as explained earlier, whereas the normalized variances of the sideward and longitudinal HBT radii show little variation with $K_T$. 
	
\begin{figure*}[!htbp]
		\centering
		\includegraphics[width=0.75\textwidth]{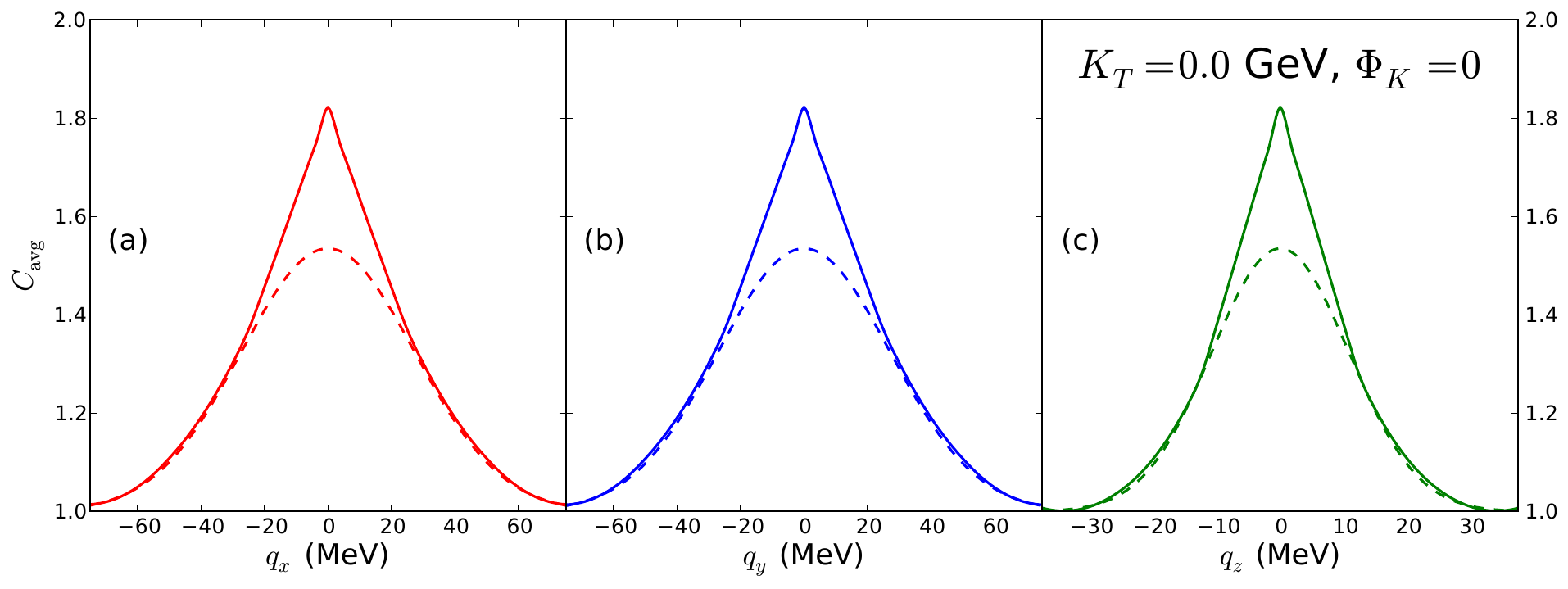}\\
		\includegraphics[width=0.75\textwidth]{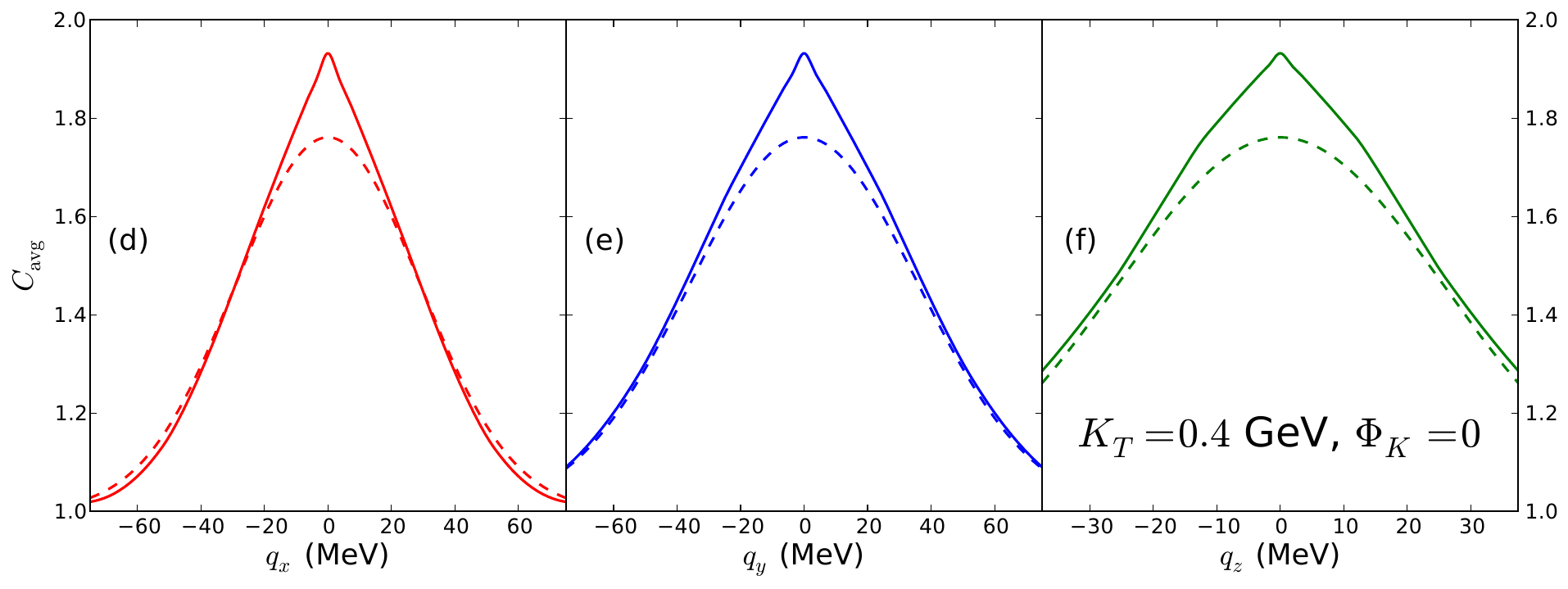}\\
		\includegraphics[width=0.75\textwidth]{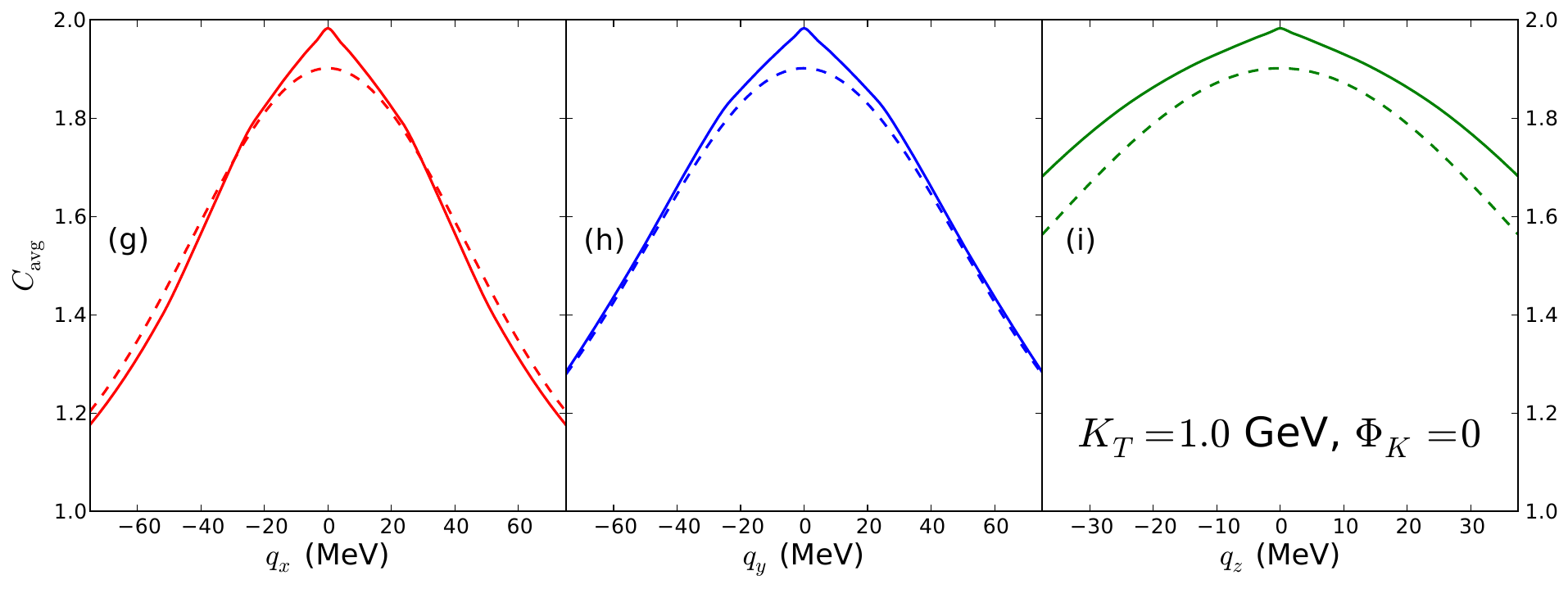}
		\caption{(Color online)
		Slices of the ensemble-averaged ($N_{\ev} = 1000$ events) correlation function 
		$C_{\mathrm{ev}}$ (solid lines), including all resonance decays, for (from top to 
		bottom) $K_T=0,\, 0.4$, and 1\,GeV, compared with the same slices of the best-fit 
		3-d Gaussian correlation function (dashed lines). These best-fit curves clearly reproduce 
		the shape of the true correlation function better at large $q$ than at small $q$.
		\label{CF_PEA_comparison}}
	\end{figure*}

	Finally we show in Fig.~\ref{CF_PEA_comparison} slices of the ensemble-averaged correlation function \eqref{corrfuncEAdefn} along the $q_x$, $q_y$, and $q_z$ axes, for three values of $K_T$ ($K_T = 0$, 0.4, and 1 GeV), together with the best three-dimensional Gaussian fit. We observe (as previously noted  \cite{Wiedemann:1996ig}) that the resonance decay effects which are most prominent at small $K_T$ (the top row) and small $q$ are not well described by the Gaussian fit function. This problem becomes less severe at larger $K_T$. The result of the poor fit near $q=0$ is a significantly reduced intercept parameter $\lambda$ extracted from the Gaussian fit than would be appropriate for describing the true value of the correlation function near $q=0$. Comparison of the solid and dashed lines allows to separate the correlation function into two contributions \cite{Csorgo:1994in}: one from the ``core'' of the emission function, describing the distribution of the directly emitted pions and those from the decay of very short-lived resonances, which dominates the large-$q$ behavior of the correlation function, and a second contribution from a ``halo'' of pions emitted by decays of long-lived resonances whose interference with ``core'' pions and with each other generate the excess of the correlation function over the Gaussian fit at small $q$ values (with $q$ components of magnitudes below 20\,MeV in our case here, corresponding to homogeneity radii of order 10\,fm).
	
	%%%%%%%%%%%%%%% Fig. 10 %%%%%%%%%%%%%%%%%%%%%
	%%%%%%%%%%%%%%%%%%%%%%%%%%%%%%%%%%%%

	%%%%%%%%%%%%%%%%%%%%%%%%%%%%%%%%%%%%%%%%%%%%%%%%%%%%%%%
	\subsection{Sensitivity of the GF HBT radii to the fit method}
	\label{Sec3d}
	%%%%%%%%%%%%%%%%%%%%%%%%%%%%%%%%%%%%%%%%%%%%%%%%%%%%%%%

\begin{figure}[!htbp]
		\centering
		\includegraphics[width=0.85\linewidth]{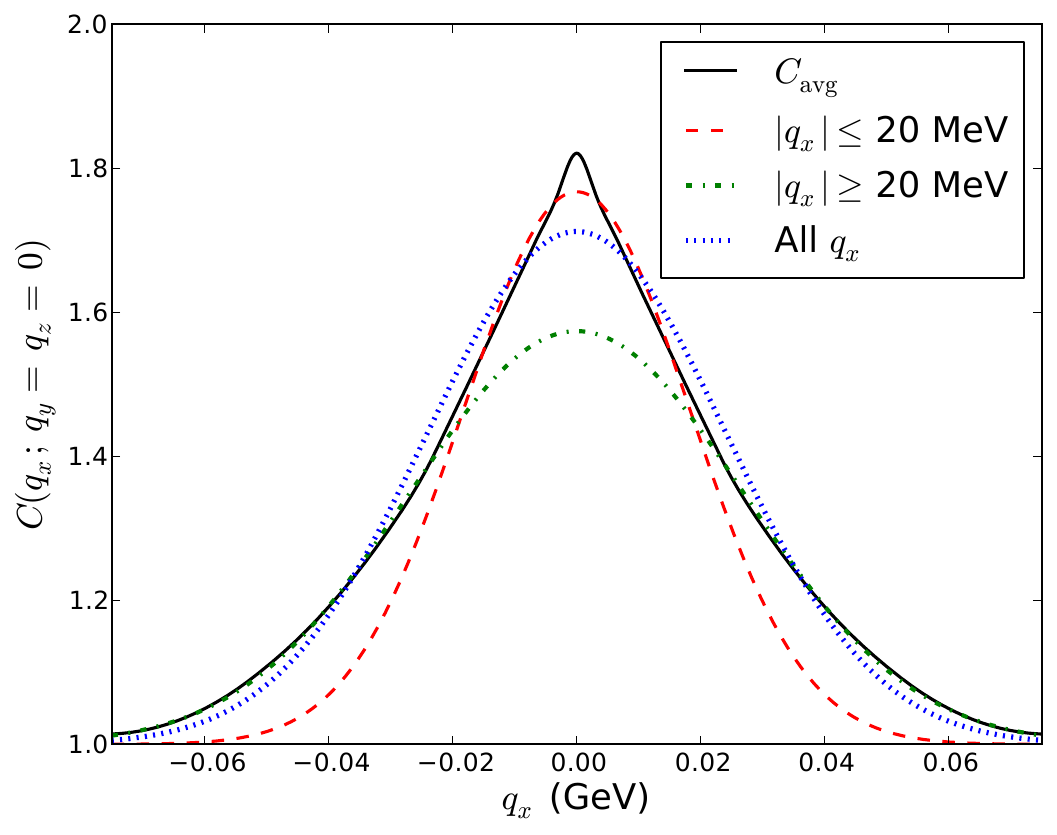}
		\caption{One-dimensional fits to $q_x$-slice of the correlation function shown in the uppermost, lefthand panel of Fig. \ref{CF_PEA_comparison}, for different ranges of $q_x$: $\l| q_x \r| \leq 20$ MeV (red, dashed curve); $\l| q_x \r| \geq 20$ MeV (green, dash-dotted curve); all $q_x$-points in the range shown (blue, dotted curve).  We see that the Gaussian best-fits depend strongly on the precise distribution of points used.
		\label{CF_fitrange_X}}
	\end{figure}

	We conclude this section by making an observation about the points chosen in the fitting process: in fitting a computed correlation function (with strongly non-Gaussian features such as those shown in Fig. \ref{CF_PEA_comparison} at small $\vec{K}$), the distribution of points used in this fit plays a significant role.  To illustrate this point, we consider in Fig. \ref{CF_fitrange_X} several one-dimensional fits to the $q_x$ slice of the correlation function plotted in the left, uppermost panel in Fig. \ref{CF_PEA_comparison}, for different choices of fit-range.  Specifically, we plot the fit curves obtained using the following sets of points:
	\begin{enumerate}
		\item All $q_x$-points with $\l| q_x \r| \leq 20$ MeV\\(red, dashed curve)
		\item All $q_x$-points with $\l| q_x \r| \geq 20$ MeV\\(green, dash-dotted curve)
		\item All $q_x$-points in the range shown\\(blue, dotted curve)
	\end{enumerate}
	%
	%%%%%%%%%%%%% Fig. 11 %%%%%%%%%%%%%%%%%
	%%%%%%%%%%%%%%%%%%%%%%%%%%%%%%%%%%%%
	We observe that these different fits vary dramatically, depending on the range and distribution of points used.  Fig. \ref{CF_fitrange_X} therefore illustrates the crucial point that theoretical HBT analyses which compute the correlation function must fit this function with the \emph{same} distribution of points used by experimentalists; otherwise, the HBT radii extracted from these Gaussian fits will not agree between theory and experiment even if the correlation functions have the same shape.
	
	%%%%%%%%%%%%%%%%%%%%%%%%%%%%%%%%%%%%%%%%%%%%%%%%%%%%%%%
	\section{Conclusions}
	\label{Sec4}
	%%%%%%%%%%%%%%%%%%%%%%%%%%%%%%%%%%%%%%%%%%%%%%%%%%%%%%%
	
	In this paper, we have presented the first calculations of the HBT radii (with resonance decay contributions) directly from Cooper-Frye integrals, for event-by-event hydrodynamics.
	
	We are certainly not the first authors to extract the HBT radii theoretically by fitting the two-particle correlation function, although previous work in this area has generally differed from the approach adopted here.  Some authors (e.g., \cite{Morita:2003mj}) formulate the correlation function in terms of Cooper-Frye integrals, as we have done here, including all relevant resonance decays, but do not perform their analysis on an event-by-event basis, meaning that they are unable to define event-by-event distributions of HBT radii.  On the other hand, some other authors \cite{Bozek:2014hwa} do compute the two-particle correlation function on an event-by-event basis, but use a statistical hadronization code and after-burner such as THERMINATOR \cite{Kisiel:2005hn} to self-consistently implement resonance decay contributions, instead of relying on the Cooper-Frye formulation to compute the resonance feeddown exactly.  What has so far \emph{not} been attempted, to the authors' knowledge, is the simultaneous incorporation of the fitted-correlator approach together with a purely hydrodynamic, Cooper-Frye formulation, with all resonances included, on an event-by-event basis.  Moreover, the resulting \emph{ensemble} of correlation functions and HBT radii has never before been studied directly, as has been done here, using both the SV and GF methods for computing the HBT radii discussed in this paper.
	
	In the case of the SV HBT radii, we find that the ensemble averaged radii with resonances are an order of magnitude larger than in the purely thermal case which has been investigated extensively elsewhere (cf. Fig. 4 of \cite{Plumberg:2015eia}).  This is a consequence of the way in which the SV radii represent the curvature of the correlator at $\vec{q}=0$, thanks to the sharp peak the correlator acquires at this point from long-lived resonances.  Once all such resonance effects are included, the qualitative features of the $R^2_{ij}$ and their $K_T$-dependence remain essentially unchanged, but the quantitative effects are drastic.	
	We also compared the event-by-event distributions of the GF HBT radii with those of the SV radii.  Interestingly, the broadening of the $R^2_o$ distribution with increasing $K_T$ appears to be a robust feature of both the SV and GF methods.  Quantitatively, the GF radii are in general smaller than their SV counterparts, once resonances are included, since the two methods differ in their representation of the global structure of the correlation function: while the SV radii represent the curvature of the correlation function at $\vec{q}=0$, the GF radii represent a best fit to the full correlation function, and therefore do not tend to overestimate the effects of long-lived resonances as severely as the SV radii do.
	
	We finally note a number of similarities and differences between our results and those presented in previous works \cite{Wiedemann:1996ig,Frodermann:2006sp}.  In particular, we note that the differences between the shapes of the correlation functions with and without resonance decays in Fig. \ref{CF_w_v_wo_res_comparison} are much larger in a genuine hydrodynamic simulation than in the hydro-motivated Gaussian source model parametrization studied in \cite{Wiedemann:1996ig}, and that these differences manifest themselves in significant quantitative effects on the extracted HBT radii (most notably, a factor of 2-3 discrepancy between the transverse radii with and without resonances, seen by contrasting the thermal radii in Refs. \cite{Plumberg:2015eia, Plumberg:2015mxa, Plumberg:2015aaw} with those shown in Fig. \ref{EAGFHBT}, which was not observed in \cite{Wiedemann:1996ig}).  Nevertheless, we find rough quantitative agreement with both \cite{Wiedemann:1996ig} and \cite{Frodermann:2006sp} in the fit radii themselves once resonance contributions are included, despite the substantial differences in the computed correlation functions.

	%%%%%%%%%%%%% Acknowledgements %%%%%%%%%%%%%%%%%%%%%%%%%%%
	\acknowledgements{
	This work was supported by the U.S.\ Department of Energy, Office of Science, Office of Nuclear
	Physics, under Grants No.~\rm{DE-SC0004286} and \rm{DE-FG02-87ER40328}, as well as (within 
	the framework of the JET Collaboration) \rm{DE-SC0004104}.  
	}
	%%%%%%%%%%%%%%%%%%%%%%%%%%%%%%%%%%%%%%%%%%%%%%%%%%%%%%%%%%

	%%%%%%%%%%%%%%%%%%%%%%%%%%%%
	%% End of main portion of paper
	%%%%%%%%%%%%%%%%%%%%%%%%%%%%

%%%%%%%% Appendices %%%%%%%%%%%%%%%%%%%%%%%%%%%%%%%%%%%%%
	
	\appendix
	%%%%%%%%%%%%%%%%%%%%%%%%%%%%%%%%%%%%%%%%%%%%%
	\section{Algorithms and approximations for efficiently computing HBT correlation functions with full resonance contributions}
	\label{App:Appendix}
	%%%%%%%%%%%%%%%%%%%%%%%%%%%%%%%%%%%%%%%%%%%%%	
	
	In this Appendix we describe the numerical code for evaluating the Fourier transformed full emission function \eqref{FTspectra} and the integration \eqref{phase_space_integrals} over the decay phase-space in particular.  We have checked that the code produces correlation functions which are in agreement with those obtained from a separate code which samples the Cooper-Frye spectra and then uses an HBT afterburner on the resulting particle pair distributions \cite{ChunMeAndUli}.  Using the simplifications and symmetries described below, the events processed in this paper required approximately 35-40 hours of CPU time each on a Intel(R) Xeon(R) X5650 2.67 GHz processor.  The grid sizes used are given below.

%%%%%%%%%%%%%%%%%%%%%%%%%%%%%%%%%%%%%%%%%%		
	\subsection{Numerical scaling}
	\label{App:A}
%%%%%%%%%%%%%%%%%%%%%%%%%%%%%%%%%%%%%%%%%%

	We begin by discussing how the full correlation function itself scales with the number of points at which the various quantities ($\vec{q}$, $\vec{K}, x^\mu$, etc.) in the calculation are evaluated.  In general, the resonance spectra are of the form
	\begin{equation}
		\int d^4 x\, e^{i q \cdot x} S(x,P),
	\end{equation}
	where is defined $q^0 = \vec{q} \cdot \vec{\beta}_{\vec{K}}$ in terms of the pair momentum $K$, while the weight $S$ is evaluated at some other momentum $P$ (one would obtain the thermal pion spectra by simply setting $P=K$).  In general, this requires a 9-dimensional grid for the evaluation of each set of weighted parent resonance spectra, with independent dimensions corresponding to differing choices of $K_T$, $\Phi_K$, $Y_K$,  $P_T$, $\Phi_P$, $Y_P$, $q_x$, $q_y$, and $q_z$.  Fortunately, however, this dimensionality can be reduced by instead treating $q^0$ as a free dimension, eliminating the dependences on $K_T$, $\Phi_K$, and $Y_K$, and only reintroducing these dependences at the end of the calculation by interpolating $q^0$ to the point that satisfies the on-shell condition \eqref{On_shell_condition_for_pions}.  For midrapidity ($Y_K = 0$) pions, which we consider exclusively in this work, this means that the Fourier-transformed spectra of each relevant particle species must be evaluated on a 7-dimensional grid, consisting of $P_T$, $\Phi_P$, $Y_P$, $q^0\equiv q_t$, $q_x$, $q_y$, and $q_z$.%
	
	For the numerical results presented in this paper, we chose the following grid sizes, unless stated otherwise:
	\begin{equation}
		N_{p_T} = 15, N_{\Phi_p} = 36, N_{q_t} = 51, N_{q_x}=N_{q_y}=N_{q_z}=7
	\end{equation}

	\subsection{Truncated and extrapolated resonance sums}
	\label{App:C}
	One technique which has proven useful for event-by-event analyses of heavy-ion collisions in the past \cite{Qiu:2012tm} requires one to sort the parent resonances by their total contributions to the (momentum-integrated) final pion yield, computing those resonances with the largest contributions first, and terminating the calculation when a fixed percentage of the total pion yield has been reached.  The authors of \cite{Qiu:2012tm} showed that using linear extrapolation to approximate the contribution to the yield from the remaining resonances offered an efficient method for obtaining estimates of the true $p_T$ spectra and anisotropic flow coefficients with all resonance contributions included.  Since the vast majority of final state resonance decay pions come from a relatively small number of parent resonances, this approach offered a significantly faster way of numerically evaluating heavy-ion observables in the context of event-by-event hydrodynamic simulations.

\begin{figure*}[!htbp]
		\centering
	\includegraphics[width=0.99\textwidth]{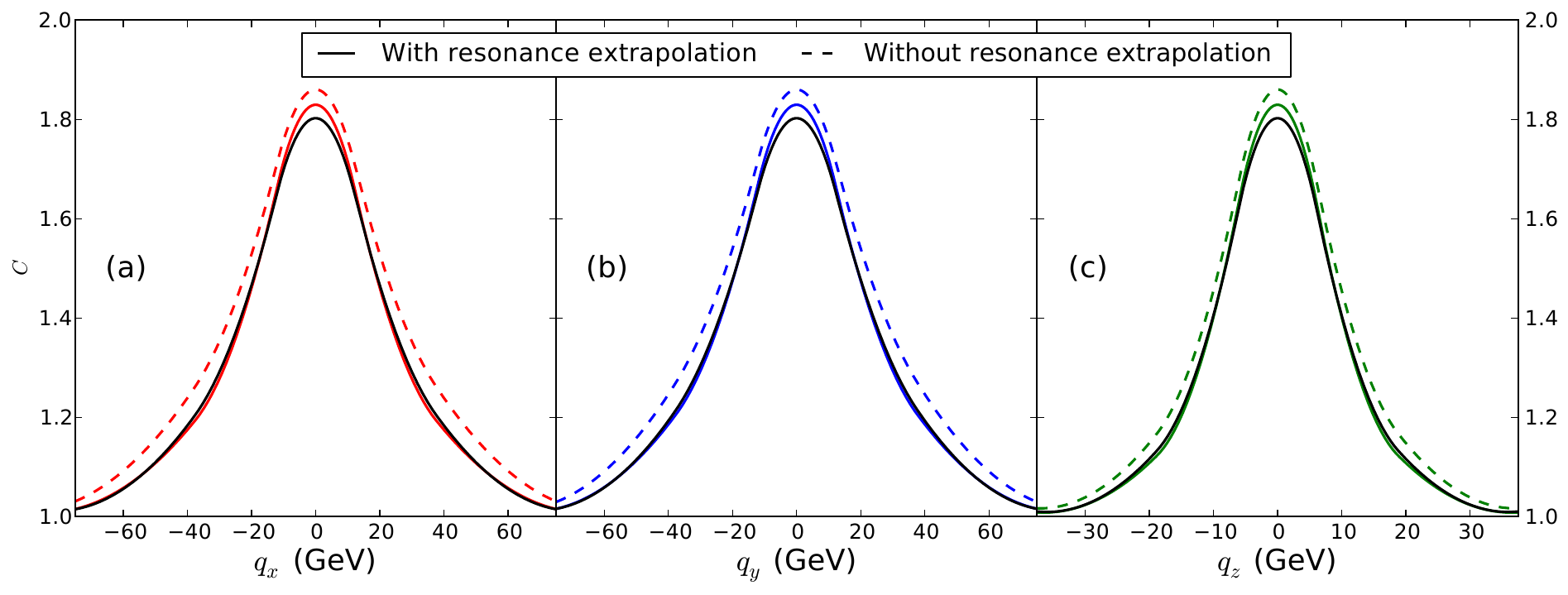}
		\caption{Three different slices ($q_x=0$, $q_y=0$, and $q_z=0$, respectively) of the correlation function at fixed $K_T$ and $\Phi_K$, comparing the truncated resonance calculation at 60\%, with (solid) and without (dashed) extrapolation over the remaining resonances, compared with the full 100\% calculation (solid black).  The correlation function was computed and extrapolated at 7 equally spaced nodes along each $q$-axis (cf.~\ref{Sec3c}) and then interpolated using a quadratic spline for aesthetic purposes.
		\label{FOintegral_resfracs_comparison}}
	\end{figure*}

	In this paper, we have adopted this same tactic for truncating and estimating the sum over parent resonances $r$ in Eq.~\eqref{resonance_decay_emission_fcn}, assuming that the rate of convergence of the Fourier-transformed decay pion spectra (as a function of resonances included) can be approximated as linear.  In Fig.~\ref{FOintegral_resfracs_comparison}, we see that this approximation works quite well: we compare the correlation function with and without extrapolation (respectively, solid and dashed colored curves) with the full result (solid black curves) which includes all resonances (and hence does not require extrapolation).  The fact that the solid black curves are mostly obscured by the solid colored curves reflects two important facts: first, that the extrapolation over the omitted resonances is necessary to capture the full correlation function using the truncated resonance sum, and second, that once this extrapolation is included, the agreement between the 60\% and 100\% curves is extremely good.  We observe that this approach works quite well for all three radii ($R^2_s$, $R^2_o$, $R^2_l$): for one-dimensional fits to the correlation functions in Fig.~\ref{FOintegral_resfracs_comparison} (with the inaccessible central point omitted), the the largest variation in the radii extracted the projected 60\% and 100\% curves was a 2.5\% change in $R^2_l$.  We conclude that resonance extrapolation provides an extremely reliable way of estimating the correlation functions and radii for all resonances relevant in heavy-ion collisions, using only a finite subset thereof.
	%%%%%%%%%%%%%%%%%%%%%%%%%%%%%%%%%%%%
	
	\subsection{Fleshing out the correlation function}
	\label{App:D}
	
	%%%%%%%%%%%%%%%%%%%%%%%%%%%%%%%%%%%%
	\begin{figure*}[!htbp]
		\centering
		\includegraphics[width=0.99\textwidth]{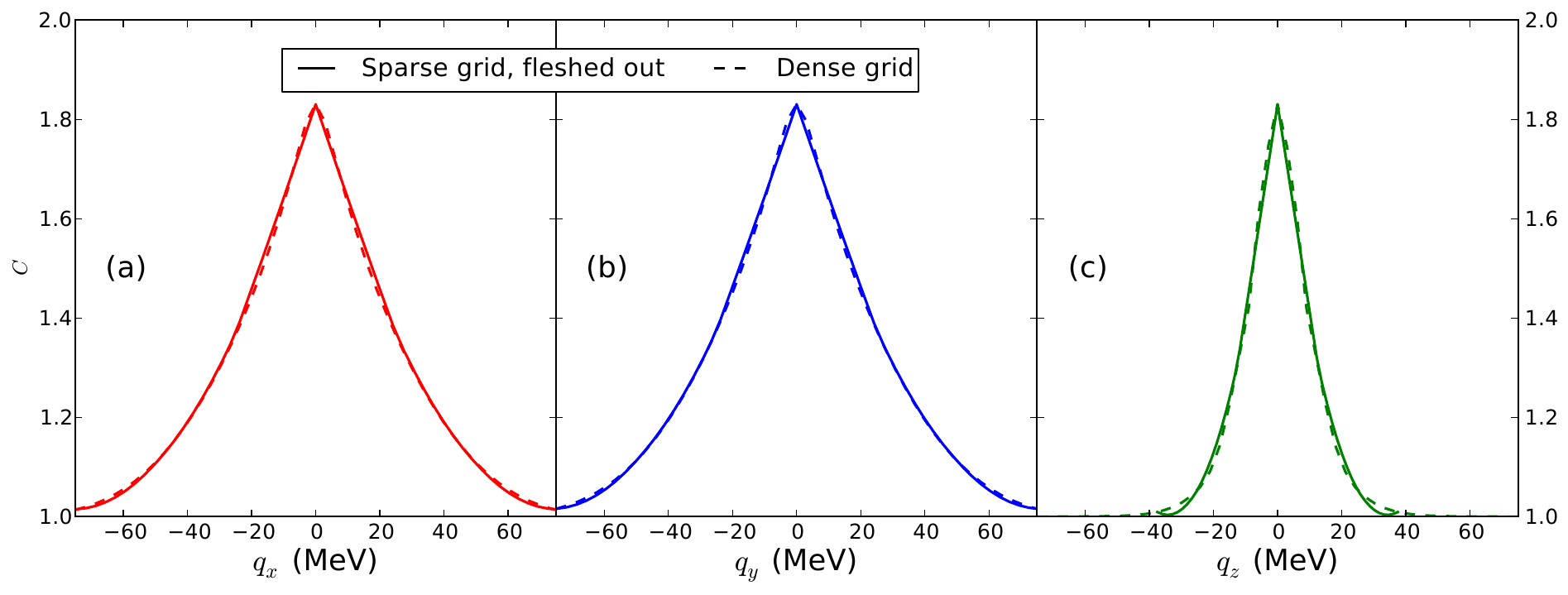}
		\caption{Three different slices ($q_x=0$, $q_y=0$, and $q_z=0$, respectively) of the correlation function at fixed $K_T$ and $\Phi_K$, illustrating our correlation function constructed by fleshing out a sparse grid (solid), compared with the same correlation function computed on a dense grid over a similar range of $q$-points (dashed).  For the first term in the numerator of \eqref{CFWR_defn}, the algorithm interpolates the logarithm of the thermal contribution linearly in $q^2$, which is an excellent approximation, since the thermal contribution is nearly Gaussian.  For the remaining two terms in the numerator of \eqref{CFWR_defn}, the algorithm uses cubic interpolation.  This approach clearly works well at all $q$, although small discrepancies emerge at large $q$ in the longitudinal ($q_z$) direction.
		\label{CF_flesh_out_comparison}}
	\end{figure*}
	%%%%%%%%%%%%%%%%%%%%%%%%%%%%%%%%%%%%

	As we have pointed out in Sec. \ref{Sec3c}, the quality of the fit to the correlation function, depends on a number of factors, including the distribution of points in $q$-space.  This is not problematic when the shape of the correlation function is Gaussian (or very nearly so).  However, when the correlation function is strongly distorted from a Gaussian shape by the inclusion of medium-lived resonances (e.g., the $\omega$ resonance, which has a width of roughly 8.5 MeV), it becomes necessary to sample the correlation function at a denser distribution of points, in order for the fitting procedure to yield well-defined and unambiguous results.  Because of the limited computational resources (as discussed above) which are typically available for performing event-by-event analyses, a sufficiently dense grid of points must be regarded as generally impractical.  What \emph{is} possible is to first compute the correlation function on a sparse grid of points, and then attempt to use this sparse grid to interpolate the correlation function to a sufficiently dense grid of points to ensure that the Gaussian fit radii become unambiguous.  We refer to this tactic as \textit{fleshing out} the correlation function, and we illustrate its effectiveness in Fig.~\ref{CF_flesh_out_comparison}.  The sparse grid of points is the same as the one defined above in Sec.~\ref{Sec3c}, consisting of 7 points in each direction.  The dense grid used here spans the same range in each direction as the sparse grid ($-75.0 \leq q_x,\,q_y \leq +75.0$ MeV and $-37.5 \leq q_z \leq 37.5$), but with 51 points in each direction.

	Clearly, although small discrepancies arise in each direction, the transverse radii are essentially identical for the curves shown in Fig.~\ref{CF_flesh_out_comparison}a and b.  The largest difference again emerges in the longitudinal ($q_z$) direction, where our difficulty at reproducing the exact correlator in the range 30 MeV $\leq \l| q_z \r| \leq$ 40 MeV leads us to overestimate $R^2_l$ by roughly 3.5\%.  The quality of the longitudinal interpolation can obviously be improved, for instance, by using a denser grid of points in this direction, at the expense of greater computational time.  We defer improvements of the ``fleshing out" technique to future studies.

	\subsection{Evaluating the Fourier-transformed spectra and resonance decays}
	\label{App:E}
	
	%%%%%%%%%%%%%%%%%%%%%%%%%%%%%%%%
	We conclude this appendix by documenting several steps which allows us to simplify and accelerate the calculation of all Fourier-transformed spectra as functions of $p_T$, $p_\phi$, and $p_Y$.  
	
	The Cooper-Frye prescription requires an integration of the distribution function over space-time coordinates, which include the space-time rapidity $\eta_s$.  We first expand the distribution function in a Boltzmann-like series of exponentials under the approximation that the particle mass $m$ is much larger than the freeze-out temperature $T$: $m/T \gg 1$.  Each term in this series expansion can then be integrated exactly over $\eta_s$, with the result expressible in terms of Bessel functions.  Below, we write down the expressions that are used to compute the thermal particle spectra: for all particles but pions, only the leading order terms in the Bessel series expansion are required.  For pions, we keep the 10 largest terms in all relevant sums, allowing us to maintain accuracy at or better than the level of $10^{-4}$.

	The Cooper-Frye integrals used in the code rely on Fourier-transforming the equilibrium distribution function with shear viscous corrections included.  The emission function and distribution function can be written in general in the form
	\begin{eqnarray}
		S(x,p) &=& \frac{1}{(2\pi)^3} \int_{\Sigma_\mathrm{f}} p{\,\cdot\,}d^3 \sigma(y)\, \delta^4 (x{-}y)\, f(y,p)\,,
	\label{App:cooper_frye_defn1}
	\\
	f(x,p) 	&=& f_0 \l(x,p\r) + \delta f \l(x,p\r) \label{App:cooper_frye_defn2}\\
			&=& \frac{1}{e^{(p \cdot u{-} \mu)/T}{-}1}{+}\frac{\chi(p^2) p^{\mu} p^{\nu} \pi_{\mu\nu} }{2 T^2 ({\cal E}{+}{\cal P})} f_0 (1{+}f_0); \nonumber
	\end{eqnarray}
	accordingly, the Fourier transform of the emission function is
	\begin{eqnarray}
		\int d^4x e^{i q \cdot x} S(x,p)
		&=& \frac{1}{(2\pi)^3} \int d\eta_s \nonumber\\
		&\times &  \int_{\Sigma_\mathrm{f}} p{\,\cdot\,}d^3\sigma(y)\,e^{i q \cdot y} f(y,p)\,.
	\end{eqnarray}
	For a system undergoing longitudinal Bj\"orken expansion, we have also that
	\begin{equation}
		p \cdot u = \gamma_\perp \l( m_\perp \cosh(p_y - \eta_s)  - \vec{p}_\perp \cdot \vec v_\perp\r)\,,
	\end{equation}
	\begin{equation}
		p{\,\cdot\,}d^3 \sigma(x) = \l( m_\perp \cosh(p_y - \eta_s)  - \vec{p}_\perp \cdot \nabla \tau_f \r) \tau_f d^2 r_\perp d\eta_s\,,
	\end{equation}
	and
	\begin{eqnarray}
		q \cdot x
			& \equiv & q^0 \tau \cosh \eta_s - \vec q_\perp \cdot \vec x_\perp - q_z \tau \sinh \eta_s \nonumber\\
			&=& \tau \l[\l( q^0 \cosh p_y - q_z \sinh p_y \r) \cosh(p_y - \eta_s) \r. \nonumber\\
			&+& \l. \l( q^0 \sinh p_y - q_z \cosh p_y \r) \sinh(p_y - \eta_s)\r] - \vec q_\perp \cdot \vec x_\perp \nonumber\\
			& \equiv & \beta \cosh \tilde \eta_s + \gamma \sinh \tilde \eta_s - \vec q_\perp \cdot \vec x_\perp,
		\label{BetaAndGammaDefinition}
	\end{eqnarray}
	where we have introduced the shorthands
	\begin{widetext}
	\iffalse
	\begin{eqnarray}
		\beta & \equiv & \tau\l( q^0 \cosh p_y - q_z \sinh p_y \r) \\
		\gamma & \equiv & \tau\l( q^0 \sinh p_y - q_z \cosh p_y \r) \\
		\tilde{\eta}_s & \equiv & p_y - \eta_s
	\end{eqnarray}
	\fi
	\begin{eqnarray}
		\beta \equiv \tau\l( q^0 \cosh p_y - q_z \sinh p_y \r) %\\
		\gamma \equiv \tau\l( q^0 \sinh p_y - q_z \cosh p_y \r) %\\
		\text{and }\tilde{\eta}_s \equiv p_y - \eta_s
	\end{eqnarray}
	Focusing on the integral over $\eta_s$, we find that the Fourier-transformed spectra can be evaluated as
	%\begin{widetext}
	%
	\begin{eqnarray}
		\int d^4x e^{i q \cdot x} S(x,p) 		& \sim & \int^{\infty}_{-\infty} d\eta_s\, \frac{ m_\perp \cosh(p_y - \eta_s)  - \vec{p}_\perp \cdot \nabla \tau_f }
												{ e^{(p \cdot u{-} \mu)/T}{\pm}1 } \l( 1 + \chi_{\mu\nu}p^\mu p^\nu\l( 1+ f_0 \r) \r)
												\nonumber\\
			&=& \int^{\infty}_{-\infty} d\tilde \eta_s\, \l( m_\perp \cosh\tilde \eta_s  - \vec{p}_\perp \cdot \nabla \tau_f \r) \exp\l( i \beta \cosh \tilde \eta_s + i \gamma \sinh \tilde \eta_s - i \vec q_\perp \cdot \vec x_\perp \r) \nonumber\\
			& \times & \sum_{k=1}^\infty {(\mp)}^{k+1}\exp\l(-\frac{k \gamma_\perp}{T}
							\l( m_\perp \cosh\tilde \eta_s  - \vec{p}_\perp \cdot \vec v_\perp - \mu\r)\r) \nonumber\\
			& \times & \l[ 1 + C \chi_{\mu\nu}p^\mu p^\nu \l( 1+ \sum_{\ell=1}^\infty {(\mp)}^{\ell+1} \exp\l(-\frac{\ell \gamma_\perp}{T}
							\l( m_\perp \cosh\tilde \eta_s  - \vec{p}_\perp \cdot \vec v_\perp - \mu\r)\r) \r) \r] \,,
		\label{longuglyequation}
	\end{eqnarray}
	\end{widetext}
	and $\chi_{\mu\nu}p^\mu p^\nu$ can be written in the form
	\begin{equation}
		\chi_{\mu\nu}p^\mu p^\nu = a \cosh^2 \tilde \eta_s + b \cosh \tilde \eta_s + c.
	\end{equation}
	The basic form of this expression is the integral
	\begin{equation}
		I_k\l( \alpha, \beta, \gamma \r)
			\equiv \int^{\infty}_{-\infty}dx\, e^{-\alpha \cosh x + i \beta \cosh x + i \gamma \sinh x} \cosh^k x\,.
	\end{equation}
	Since it is obvious that
	\begin{equation}
		I_k\l( \alpha, \beta, \gamma \r) = \l( - \frac{d}{d\alpha} \r)^k I_0\l( \alpha, \beta, \gamma \r)\,,
	\end{equation}
	we only need to compute $I_0\l( \alpha, \beta, \gamma \r)$.  Making the change of variable $u = \sinh x$, $du = \cosh x \,dx = \sqrt{u^2+1} dx$, we find that
	\begin{eqnarray}
		I_0\l( \alpha, \beta, \gamma \r)
			 &=& \int^{\infty}_{-\infty}\frac{du}{\sqrt{u^2+1}} e^{-\l( \alpha - i \beta \r) \sqrt{u^2+1} + i \gamma u} \nonumber\\
			 &=& 2 \int^{\infty}_{0}\frac{du\,\cos\l( \gamma u \r)}{\sqrt{u^2+1}} e^{-\l( \alpha - i \beta \r) \sqrt{u^2+1}} \nonumber\\
			 &=& 2 K_0\l( \sqrt{\l( \alpha - i \beta \r)^2 + \gamma^2} \r).
	\end{eqnarray}
	\begin{widetext}
	Defining
	%
	\iffalse
	\begin{eqnarray}
		A &\equiv & e^{- i \vec q_\perp \cdot \vec x_\perp} m_\perp, \, \\
		B &\equiv & e^{- i \vec q_\perp \cdot \vec x_\perp}\vec{p}_\perp \cdot \nabla \tau_f, \, \\
		\alpha &\equiv & \frac{\gamma_\perp m_\perp}{T}, \, \\
		f_\perp &\equiv & \exp\l(\frac{\gamma_\perp}{T}
							\l(\vec{p}_\perp \cdot \vec v_\perp + \mu\r)\r)\,,
	\end{eqnarray}
	\fi
	\begin{equation}
		A \equiv e^{- i \vec q_\perp \cdot \vec x_\perp} m_\perp, \, %\\
		B \equiv e^{- i \vec q_\perp \cdot \vec x_\perp}\vec{p}_\perp \cdot \nabla \tau_f, \, %\\
		\alpha \equiv \frac{\gamma_\perp m_\perp}{T}, \, %\\
		\text{and }f_\perp \equiv \exp\l(\frac{\gamma_\perp}{T}
							\l(\vec{p}_\perp \cdot \vec v_\perp + \mu\r)\r)\,,
	\end{equation}
	we can compute the Bessel series expansion of \eqref{longuglyequation} term-by-term to obtain
	%
	%\begin{widetext}
	\begin{eqnarray}
		\int d^4x e^{i q \cdot x} S(x,p)
			& \sim & \int^{\infty}_{-\infty} d\tilde \eta_s\, \l( A \cosh\tilde \eta_s  - B \r) \exp\l( i \beta \cosh \tilde \eta_s + i \gamma \sinh \tilde \eta_s \r)\nonumber\\
			& \times & \sum_{k=1}^\infty \mp{(\mp f_\perp)}^{k}\exp\l(-k \alpha \cosh\tilde \eta_s \r) \l[ 1 + C \chi_{\mu\nu}p^\mu p^\nu \l( 1+ \sum_{\ell=1}^\infty \mp{(\mp f_\perp)}^{\ell} \exp\l(- \ell \alpha \cosh\tilde \eta_s\r) \r) \r] \nonumber\\
			&=& \int^{\infty}_{-\infty} dx\, \l( A \cosh x  - B \r) \sum_{k=1}^\infty \mp{(\mp f_\perp)}^{k}\exp\l(-k \alpha \cosh x + i \beta \cosh x + i \gamma \sinh x \r)\nonumber\\
			& \times & \l[ 1 + C \l( a \cosh^2 x + b \cosh x + c \r) \l( 1+ \sum_{\ell=1}^\infty \mp{(\mp f_\perp)}^{\ell} \exp\l(- \ell \alpha \cosh x\r) \r) \r] %\\
	\end{eqnarray}
	\begin{eqnarray}
			&=& \sum_{k=1}^\infty \mp{(\mp f_\perp)}^{k}\l( A I_1(k \alpha, \beta, \gamma) - B I_0(k \alpha, \beta, \gamma) \r) \nonumber\\
			&+& C \sum_{k=1}^\infty \mp{(\mp f_\perp)}^{k} \l( A a I_3(k \alpha, \beta, \gamma) + \l( a B + b A \r)I_2(k \alpha, \beta, \gamma) \r. \nonumber\\
			&& \l. + \l( b B + c A \r)I_1(k \alpha, \beta, \gamma) + c B I_0(k \alpha, \beta, \gamma) \r) \nonumber\\
			& \mp & C \sum_{k=1}^\infty \sum_{\ell=1}^\infty {(\mp f_\perp)}^{k+\ell} \l( A a I_3((k+\ell) \alpha, \beta, \gamma) + \l( a B + b A \r)I_2((k+\ell) \alpha, \beta, \gamma) \r. \nonumber\\
			&& \l. + \l( b B + c A \r)I_1((k+\ell) \alpha, \beta, \gamma) + c B I_0((k+\ell) \alpha, \beta, \gamma) \r)\,. \nonumber\\
			&&
	\end{eqnarray}
	\end{widetext}
	Fortunately, the Boltzmann approximation (i.e., keeping only the first term in each of these sums) works already extremely well for all hadrons other than pions.  In the Boltzmann limit, the above result simplifies to
	\begin{eqnarray}
		&& \!\!\!\!\!\!\!\!\! \int d^4x e^{i q \cdot x} S(x,p) \nonumber\\
			& \sim & f_\perp\l( A I_1(\alpha, \beta, \gamma) - B I_0(\alpha, \beta, \gamma) \r) \nonumber\\
			&+& C f_\perp \l( A a I_3(\alpha, \beta, \gamma) + \l( a B + b A \r)I_2(\alpha, \beta, \gamma) \r. \nonumber\\
			&& \l. + \l( b B + c A \r)I_1(\alpha, \beta, \gamma) + c B I_0(\alpha, \beta, \gamma) \r) \nonumber\\
			& \mp & C f_\perp^2 \l( A a I_3(2 \alpha, \beta, \gamma) + \l( a B + b A \r)I_2(2 \alpha, \beta, \gamma) \r. \nonumber\\
			&& \l. + \l( b B + c A \r)I_1(2 \alpha, \beta, \gamma) + c B I_0(2 \alpha, \beta, \gamma) \r)\,, \nonumber\\
	\end{eqnarray}
	where finally, with $z \equiv \sqrt{(\alpha - i \beta)^2 + \gamma^2}$,
	\begin{eqnarray}
	I_0(\alpha, \beta, \gamma) &=& 2 K_0\l(z\r)\,, \nonumber\\
	I_1(\alpha, \beta, \gamma) &=& \frac{2 (\alpha - i \beta) K_1\l(z\r)}{z}\,, \nonumber\\
	I_2(\alpha, \beta, \gamma) &=& \frac{2 (\alpha - i \beta)^2 K_0\l(z\r)}{z^2} + \frac{2 (z^2-2\gamma^2) K_1\l(z\r)}{z^3}\,, \nonumber\\
	I_3(\alpha, \beta, \gamma) &=& \frac{2}{z^5} (\alpha - i \beta)
		\l[z\l( z^2-4\gamma^2 \r) K_0\l(z\r) \r.\nonumber\\
		&& \l. + (2z^2+z^4-\gamma^2z^2+8\gamma^2) K_1\l(z\r)\r]\,. \nonumber\\
	\label{IkFunctions}
	\end{eqnarray}
	Inspection of these results reveals that they are even in $\gamma$, while their real (imaginary) parts are even (odd) in $\beta$.  This will be useful below.

	%%%%%%%%%%%%%%%%%%%%%%%%%%%%%%%%%%%%
	In addition to performing the $\eta_s$-integrals analytically as detailed above, the full calculation (of all thermal resonance spectra and subsequent resonance feeddown) we identify and exploit several symmetries which can be used to shorten and/or accelerate the calculation of the correlation function.  There are three symmetries which are useful for our purposes here; they are:
	\begin{enumerate}
		\item \textit{Symmetry under $q \to -q$.}  This symmetry follows trivially by replacing $e^{i q \cdot x}$ in all Fourier integrals with $\cos(q \cdot x) + i \sin (q \cdot x)$, and noting that the first term is even under this symmetry, while the second term is odd.  Consequently, the Fourier moments only need to be calculated for half of $q$-space (say, $q^0 \geq 0$) and then reflected to the other half (with the odd moments receiving an additional minus sign upon reflection).
		\item \textit{Symmetry under $q_z \to -q_z$ and $y \to -y$ simultaneously.}  This symmetry follows by noting that $q_z$ and $y$ enter into the Fourier moments only in the combinations $\beta$ and $\gamma$ given in Eq.~\eqref{longuglyequation}.  The Fourier moments, in turn, depend only on the functions Eqs.~\eqref{IkFunctions}, which depend only on $\beta$ and $\gamma^2$.  Taking $q_z \to -q_z$ and $y \to -y$ simultaneously thus takes $\beta \to \beta$ and $\gamma \to -\gamma$, leaving the Fourier moments unchanged.  This symmetry can be exploited by computing the full $y$ dependence and half the $q_z$ dependence of the moments, and reflecting to the other half as above.
		\item \textit{Symmetry under reflection of $y$ about $y_{\mathrm{sym}}$ and $\vec{q}_\perp \to \pm\vec{q}_\perp$ simultaneously}.  The third symmetry also arises by studying the structure of $\beta$ and $\gamma$.  We first define $y_{\mathrm{sym}}$ to be
		\begin{equation}
			y_{\mathrm{sym}} \equiv \frac{1}{2} \log \l| \frac{q^0+q_z}{q^0-q_z} \r|\,.
		\end{equation}
		Then a little algebra reveals that taking $y \to \bar y \equiv 2 y_{\mathrm{sym}} - y$ is a convenient reflection point for the Fourier moments.  Specifically, one can show that
		\begin{eqnarray}
			\beta\l( \bar{y} \r)
				&=& \mathrm{sgn}\l( \l( q^0 \r)^2-q_z^2 \r) \beta\l(  y \r)\,, \\
			\gamma\l( \bar{y} \r)
				&=& -\mathrm{sgn}\l( \l( q^0 \r)^2-q_z^2 \r) \gamma\l(  y \r)\,.
		\end{eqnarray}
		Moreover, note that the reflection point $y_{\rm sym}$ is the same for all resonances, since it depends only on $q^0$ and $q_z$.  Additionally, since the Fourier moments depend only on $\gamma^2$, reflection of $y \to \bar y$ is a symmetry of the Fourier moments only when $\l| q^0 \r| \geq \l| q_z \r|$; the case where $\l| q^0 \r| \leq \l| q_z \r|$ will be discussed in greater detail below.
		
		First, we describe how the Fourier moment calculation is organized.  We split the generic Fourier integral of the emission function for a resonance $r$ into different terms for convenience:
		\begin{widetext}
		\begin{eqnarray}
			\int_x e^{i q \cdot x} S_r(x,K)
				&=& \int_x \l( \cos \phi_T + i \sin \phi_T \r )\l( \cos \phi_L + i \sin \phi_L \r) S_r(x,K) \\
				&=& \int_x \cos \phi_L \cos \phi_T S_r(x,K) - i \int_x \cos \phi_L \sin \phi_T S_r(x,K) \nonumber\\
				&+& i \int_x \sin \phi_L \cos \phi_T S_r(x,K) + \int_x \sin \phi_L \sin \phi_T S_r(x,K) \\
				& \equiv & S^{CC}_r + i S^{CS}_r + i S^{SC}_r + S^{SS}_r
			\label{FourierMomentsDefinitions}
		\end{eqnarray}
		\end{widetext}
		where $\phi_T \equiv \vec{q}_\perp \cdot \vec{x}_\perp$, $\phi_L \equiv q^0 t - q_z z$. Also, in the first index, $C$ and $S$ label the cosine and sine components in $\phi_L$, while in the second index, $C$ and $S$ label the cosine and sine part in $\phi_L$.  As we shall see, the longitudinal $C$ and $S$ moments each have a definite parity under $y \to \bar y$.  Thus, each of these Fourier moments of the source function for resonance $r$ is computed independently.  To see how each these moments is related to the moments of its daughter particles, we substitute \eqref{FourierMomentsDefinitions} into the phase-space integrals which yield the contributions to the daughter moments:
		\begin{widetext}
		\begin{eqnarray}
			S^{CC}_{r \to r'} + i S^{CS}_{r \to r'} + i S^{SC}_{r \to r'} + S^{SS}_{r \to r'}
				&=& \sum_{k=\pm} \int_{\mathbf R} \frac{1+i \alpha^k}{1+\l( \alpha^k \r)^2} \l( S^{CC}_r + i S^{CS}_r + i S^{SC}_r + S^{SS}_r \r) \nonumber\\
				&=& \sum_{k=\pm} \int_{\mathbf R} \l( 1+\l( \alpha^k \r)^2 \r)^{-1} \l[ \l( S^{CC}_r - \alpha^k S^{CS}_r \r) + i \l( S^{CS}_r + \alpha^k S^{CC}_r \r) \r. \nonumber\\
				&& \qquad \qquad \qquad \qquad \qquad \l. + i \l( S^{SC}_r + \alpha^k S^{SS}_r \r) + \l( S^{SS}_r - \alpha^k S^{SC}_r \r) \r]\,.
			\label{FMthruPhaseSpaceIntegrals}
		\end{eqnarray}
		where $\alpha^k \equiv q \cdot P^k / (M \Gamma)$ and 
		\begin{equation}
			\int_{\mathbf{R}} \equiv M \int^{s_+}_{s_-} ds\,g(s) 
				\int^{+1}_{-1}\frac{\Delta Y dv}{\sqrt{m_{\perp}^2 \cosh^2(v \Delta Y) - p_{\perp}^2}} \int^{\pi}_0 
				d\zeta \, \l( \overline{M}_{\perp} + \Delta M_{\perp} \cos \zeta \r) .
		\label{App:phase_space_integrals}
		\end{equation}
		\end{widetext}			
		Each of the terms on the lefthand side of \eqref{FMthruPhaseSpaceIntegrals} should be identified with the respective term on the righthand size in the square brackets.  Note also that $\overline{M}_{\perp}$ and $\Delta M_{\perp}$ are even functions of $v \Delta Y$.
		
		We need to know whether the phase-space integrals \eqref{FMthruPhaseSpaceIntegrals} and \eqref{App:phase_space_integrals} respect the reflection symmetries of the Fourier moments of the parent resonance (so that daughter particles inherit the same set of reflection symmetries).  To show this is just a few lines: using the shorthand
		\begin{equation}
			\tilde S(q, K) \equiv \int d^4x e^{i q \cdot x} S(x,K)
		\end{equation}
		to abbreviate the Fourier transform, we can write
		\begin{widetext}
		\begin{eqnarray}
		\tilde{S}_{r \to r'}(q; m, \vec{p}_\perp, \bar y)
				&=& M  \sum_{k=\pm} \int^{s_+}_{s_-} ds\,g(s) 
				\int^{+1}_{-1}\frac{\Delta Y dv}{\sqrt{m_{\perp}^2 \cosh^2(v \Delta Y) - p_{\perp}^2}} \int^{\pi}_0 
				d\zeta \, \l( \overline{M}_{\perp} + \Delta M_{\perp} \cos \zeta \r) \nonumber\\
				& \times & \frac{1}{1-i \bar \alpha^k} \int_x e^{i q \cdot x} S_{r \to r'}(x; M, \vec{P}^k_\perp, 2Y_{\rm sym}-y+v \Delta Y) %\nonumber\\
		\end{eqnarray}
		\begin{eqnarray}
				&\eqcom{v \to -v}& M  \sum_{k=\pm} \int^{s_+}_{s_-} ds\,g(s) 
				\int^{-1}_{+1}\frac{-\Delta Y dv}{\sqrt{m_{\perp}^2 \cosh^2(-v \Delta Y) - p_{\perp}^2}} \int^{\pi}_0 
				d\zeta \, \l( \overline{M}_{\perp} + \Delta M_{\perp} \cos \zeta \r) \nonumber\\
		%\end{eqnarray}
		%\begin{eqnarray}
				& \times & \frac{1}{1-i \bar \alpha^k} \int_x e^{i q \cdot x} S_{r \to r'}(x; M, \vec{P}^k_\perp, 2Y_{\rm sym}-y-v \Delta Y) \nonumber\\
				&=& \sum_{k=\pm} \int_{\mathbf{R}} \frac{1+i \bar \alpha^k}{1 + \l(\bar \alpha^k\r)^2} \int_x e^{i q \cdot x} S_{r \to r'}(x; M, \vec{P}^\pm_\perp, \bar Y)\,,
		\end{eqnarray}
		\end{widetext}
		where the barred quantities have been reflected about $y_{\mathrm{sym}}$, and
		\begin{eqnarray}
			\bar \alpha^k
				& \equiv & \alpha^k\l(\bar{Y}\r) \\
				&=& \frac{M_\perp}{M \Gamma} \mathrm{sgn}\l( \l( q^0 \r)^2-q_z^2 \r) \beta\l( Y \r) - \frac{\vec{q}_\perp \cdot \vec{P}^k_\perp}{M \Gamma}\,.\nonumber\\
		\end{eqnarray}
		To proceed further, we now consider the cases $\l| q^0 \r| \geq \l| q_z \r|$ and $\l| q^0 \r| < \l| q_z \r|$ separately.
		\begin{enumerate}
			\item $\l| q^0 \r| \geq \l| q_z \r|$.  In this case, $\beta \to \beta$ and $\gamma^2 \to \gamma^2$ when $y \to \bar y$, so that $S^{CC}_r$, $S^{CS}_r$, $S^{SS}_r$, and $S^{SC}_r$ are all symmetric; i.e.,
			\begin{eqnarray}
				S^{CC}_r\l( \vec{q}_\perp; \bar Y \r) &=& S^{CC}_r\l( \vec{q}_\perp; Y \r) \\
				S^{CS}_r\l( \vec{q}_\perp; \bar Y \r) &=& S^{CS}_r\l( \vec{q}_\perp; Y \r) \\
				S^{SC}_r\l( \vec{q}_\perp; \bar Y \r) &=& S^{SC}_r\l( \vec{q}_\perp; Y \r) \\
				S^{SS}_r\l( \vec{q}_\perp; \bar Y \r) &=& S^{SS}_r\l( \vec{q}_\perp; Y \r)\,.
			\end{eqnarray}
			Furthermore, it is obvious that $\bar \alpha^k \to \alpha^k = \l( M_\perp \beta\l( Y \r) - \vec{q}_\perp \cdot \vec{P}^k_\perp\r)/(M \Gamma)$ as well.  Together, with \eqref{FMthruPhaseSpaceIntegrals}, this implies that
			\begin{eqnarray}
				S^{CC}_{r \to r'}\l( \vec{q}_\perp; \bar y \r) &=& S^{CC}_{r \to r'}\l( \vec{q}_\perp; y \r) \\
				S^{CS}_{r \to r'}\l( \vec{q}_\perp; \bar y \r) &=& S^{CS}_{r \to r'}\l( \vec{q}_\perp; y \r) \\
				S^{SC}_{r \to r'}\l( \vec{q}_\perp; \bar y \r) &=& S^{SC}_{r \to r'}\l( \vec{q}_\perp; y \r) \\
				S^{SS}_{r \to r'}\l( \vec{q}_\perp; \bar y \r) &=& S^{SS}_{r \to r'}\l( \vec{q}_\perp; y \r)\,,
			\end{eqnarray}
			i.e., the symmetries of each Fourier moment are preserved by the phase-space integrals.
			\item $\l| q^0 \r| < \l| q_z \r|$.  In this case, reflection about $y_{\mathrm{sym}}$ by itself is not a symmetry of the Fourier moments after the phase space integration, since now
			\begin{eqnarray}
				\bar \alpha^k
					&=& \frac{ - M_\perp \beta\l( Y \r) - \vec{q}_\perp \cdot \vec{P}^k_\perp}{M \Gamma} \nonumber\\
					&\neq & \alpha^k\,.
			\end{eqnarray}
			Fortunately, however, some symmetry can be restored if the reflection about $y_{\mathrm{sym}}$ is accompanied in this case by the reflection $\vec{q}_\perp \to -\vec{q}_\perp$, in which case
			\begin{eqnarray}
				\bar \alpha^k
					&=& - \frac{ M_\perp \beta\l( Y \r) - \vec{q}_\perp \cdot \vec{P}^k_\perp}{M \Gamma} \nonumber\\
					&=& - \alpha^k\,.
			\end{eqnarray}
			Under this combined transformation, we now find that
			\begin{eqnarray}
				S^{CC}_r\l( -\vec{q}_\perp; \bar y \r) &=& S^{CC}_r\l( \vec{q}_\perp; y \r) \\
				S^{CS}_r\l( -\vec{q}_\perp; \bar y \r) &=& - S^{CS}_r\l( \vec{q}_\perp; y \r) \\
				S^{SC}_r\l( -\vec{q}_\perp; \bar y \r) &=& - S^{SC}_r\l( \vec{q}_\perp; y \r) \\
				S^{SS}_r\l( -\vec{q}_\perp; \bar y \r) &=& S^{SS}_r\l( \vec{q}_\perp; y \r)\,,
			\end{eqnarray}
			as follows immediately from Eq.~\eqref{FourierMomentsDefinitions}.  Taking all of this together with \eqref{FMthruPhaseSpaceIntegrals}, one finds that
			\begin{eqnarray}
				S^{CC}_{r \to r'}\l( -\vec{q}_\perp; \bar y \r) &=& S^{CC}_{r \to r'}\l( \vec{q}_\perp; y \r) \\
				S^{CS}_{r \to r'}\l( -\vec{q}_\perp; \bar y \r) &=& -S^{CS}_{r \to r'}\l( \vec{q}_\perp; y \r) \\
				S^{SC}_{r \to r'}\l( -\vec{q}_\perp; \bar y \r) &=& -S^{SC}_{r \to r'}\l( \vec{q}_\perp; y \r) \\
				S^{SS}_{r \to r'}\l( -\vec{q}_\perp; \bar y \r) &=& S^{SS}_{r \to r'}\l( \vec{q}_\perp; y \r)\,,
			\end{eqnarray}
			so that the symmetries are again preserved in this case.
		\end{enumerate}
		\end{enumerate}
	
%%%%%%%%%%%%%%% Bibliography %%%%%%%%%%%%%%%%%%%%%%%%%%%%

	%% End bibliography

	%% End of document
\end{document}